\documentclass[
preprint,
aps,prd,
nofootinbib,
superscriptaddress,
showpacs,
tightenlines,
]{revtex4}
\usepackage{amsmath}
\usepackage{amssymb}
\usepackage{bm}
\usepackage[dvips]{color,graphicx}

\begin{document}

\title{Nucleon Tensor Charge from Exclusive $\pi^o$ Electroproduction}  
\author{Saeed Ahmad} 
\email{sl4y@virginia.edu}
\affiliation{Department of Physics, University of Virginia, Charlottesville, VA 22901, USA.}

\author{Gary R.~Goldstein} 
\email{gary.goldstein@tufts.edu}
\affiliation{Department of Physics and Astronomy, Tufts University, Medford, MA 02155 USA.}
 
\author{Simonetta Liuti} 
\email{sl4y@virginia.edu}
\affiliation{Department of Physics, University of Virginia, Charlottesville, VA 22901, USA.}

\begin{abstract}
Exclusive $\pi^o$ electroproduction from nucleons is suggested  
for extracting the tensor charge and other quantities related to transversity
from experimental data.   
This process isolates C-parity odd and chiral odd combinations of t-channel exchange 
quantum numbers. In a hadronic picture it connects the meson production amplitudes to C-odd Regge
exchanges with final state interactions. In a description based on partonic degrees of freedom,
the helicity structure for this C-odd process relates to the quark helicity flip, 
or chiral odd generalized parton distributions.
This differs markedly from deeply virtual Compton scattering, and both vector meson 
and charged $\pi$ electroproduction, where the axial charge can enter the amplitudes.
Contrarily the tensor charge enters the $\pi^o$ process. The connection through the helicity description
of the process to both the partonic and hadronic perspectives is studied and exploited in model
calculations to indicate how the tensor charge and other transversity parameters can be related to cross
section and spin asymmetry measurements over a broad range of kinematics.     
\end{abstract}

\maketitle
\baselineskip 3.0ex

\section{Introduction}
\begin{figure}
\hspace{-0.5cm}
{\bf (a)}
\includegraphics[width=11.1cm]{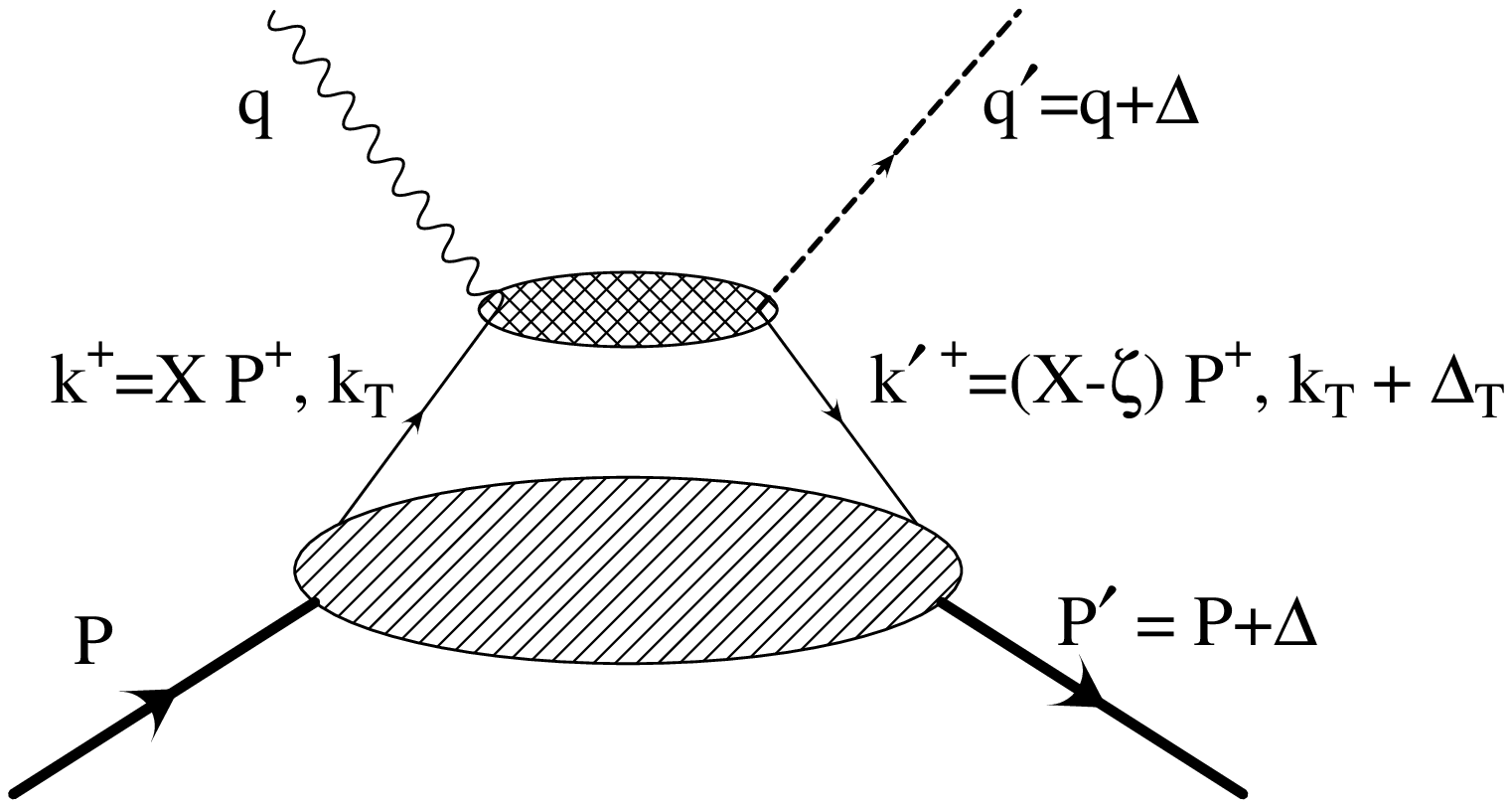}
{\bf (b)} 
\includegraphics[width=3.5cm]{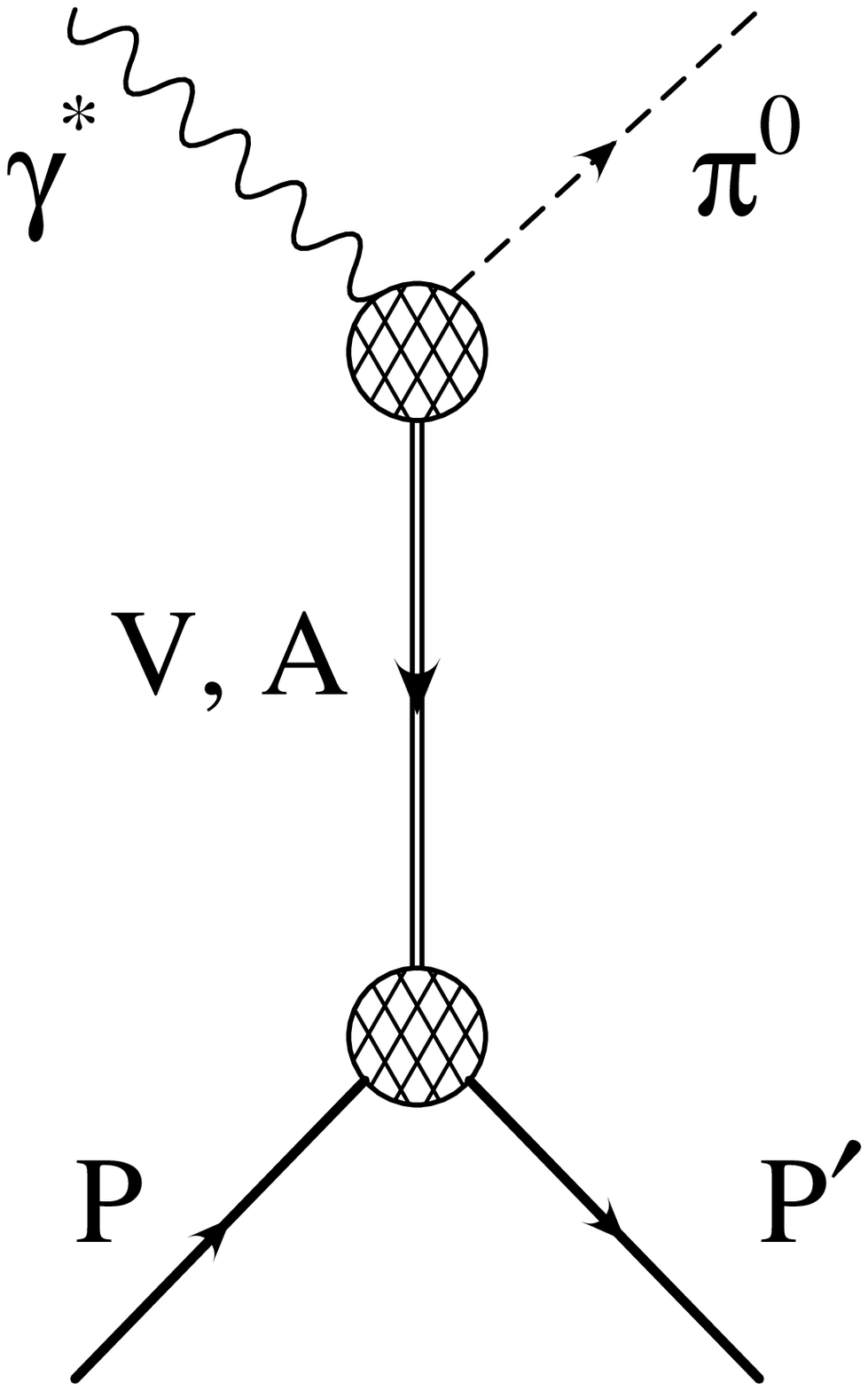}
\caption{$\pi^o$ electroproduction. Left:  partonic 
degrees of freedom interpretation; Right: $t$-channel exchange diagram.} 
\label{fig1} 
\end{figure}

A dynamical 
mechanism for the process $\gamma^* P \rightarrow \pi^0 P^\prime$ 
is proposed  
that allows for the  extraction of the tensor charge from experiment. The basis for this 
approach
is in the relation 
between a hadronic description of the process, in terms of Regge poles and cuts, and the partonic 
description, in terms 
of Generalized Parton Distributions (GPDs) \cite{DMul1,Ji1,Rad1}. The latter provides a formal 
connection to the 
transversity distribution of the nucleon, $h_1$, and the helicity amplitudes that are key to 
parameterizing the hadronic 
description. In the following we will use this connection as a guide in exploring the observables 
that isolate the 
transversity. 
The two models, GPD (Fig.\ref{fig1}a) and Regge (Fig.\ref{fig1}b), will illustrate how the 
extraction of transversity and 
the tensor charge can proceed experimentally.  
A key point in our approach is that  deeply virtual  
$\pi^o$ (as well as $\eta, \eta^\prime$) production off a proton target 
is clearly distinct from the other types of meson
production processes in that it involves the transition of a (virtual) photon with 
$J^{PC}=1^{--}$ to a $J^{PC}=0^{-+}$ state ({\it i.e.} the final $\pi^o$ or $\eta$, $\eta^\prime$)
requiring 
odd $C-$parity and chiral odd  t-channel quantum numbers.
As a consequence,  in a partonic description such as the one depicted in Fig.{\ref{fig1}a}, 
the "outgoing" and "returning" quark helicities need to be opposite to one another. 
A similar picture can be obtained in the Regge model as dictated by duality.
Therefore, $\pi^o$ and $\eta$, $\eta^\prime$  electroproduction off a proton single out chiral-odd structures 
of the target. 
Another important consequence is that the collinear, leading twist $\gamma^\mu \gamma^5$  type 
contribution to the $\pi^o$ wave function does not have the correct chirality
for  the electroproduction process. Consideration of Orbital Angular Momentum (OAM)
in the wave function allowed us however  to overcome this problem, as we will explain below.

GPDs are ``off-forward'' contributions, 
that allow access to partonic configurations 
with a given longitudinal momentum fraction, similarly to Deep Inelastic Scattering (DIS) 
processes, but also at a specific (transverse) location inside the hadron \cite{Bur}. 
They parameterize the nucleon 
vertex in the process depicted in Fig.~\ref{fig1}a in terms of three 
kinematical invariants, besides
the initial photon's virtuality, $Q^2$: the longitudinal momentum transfer,
$\zeta = Q^2/2(Pq)$, the four-momentum transfer squared, $\Delta^2=-t$, and the
variable $X=(kq)/(Pq)$, representing the Light Cone (LC) momentum fraction carried by the struck 
parton with momentum $k$ (see \cite{Die_rev,BelRad,BofPas}) for reviews). 
\footnote{
The relations between the variables used in this paper and the analogous set of kinematical variables 
in the  ``symmetric'' system, frequently used in the literature are given along with 
the definitions of the hadronic tensors components in Refs.~\cite{Die_rev,BelRad}.}  

As initially pointed out in Ref.\cite{Diehl_01} one can construct four quark helicity flip
distributions: $H_T^q$, $E_T^q$, $\tilde{H}_T^q$, $\tilde{E}_T^q$, representing a complete
set. 
\footnote{A similar argument as in Ref.\cite{Diehl_01} can be extended to define a corresponding
number of gluon helicity flip distributions that, however, will not enter our discussion of 
transversity.}
In particular $H_T^q$,   
which need not vanish in the forward limit, is related to transversity through the following properties:
\begin{eqnarray}
\label{transversity_def}
\int\limits_{\zeta-1}^1 H_T^q(X,\zeta,t) \, dX = A_{T10}^q(t) \\
\label{transv_forward}
H_T^q(X,0,0) = h_1^q(X)  
\end{eqnarray}
The form factor $A_{T10}(t)$ gives the tensor 
charge for $t \rightarrow 0$, $A_{T10}(0) \equiv \delta q$. 
$h_1(X)$ is the transversity structure function~\cite{JafJi}.
Eqs.(\ref{transversity_def}) and (\ref{transv_forward}) are analogous relations to the ones for the twist two, chiral-even, 
unpolarized and longitudinally polarized distributions cases.  

A substantial amount of literature exists on the connection between Transverse Momentum Distributions 
(TMDs) and GPDs \cite{Bur,Bur2,Bur3,DieHag,Metz}.
TMDs are the soft matrix elements in deep inelastic semi-inclusive processes and therefore
they are by definition forward quantities.  
The non-trivial role played by Initial State Interactions (ISI) and Final State Interactions (FSI) 
allows one however to access features of the motion of partons in the transverse direction. In particular, in 
Ref.\cite{Bur2} it was proposed that the transverse momentum asymmetry of 
the final quarks -- generating the Sivers function -- 
can be related to the transverse spatial asymmetry, through a chromodynamic lensing effect.   
Notwithstanding the appeal of the physical ideas connecting transversity and transverse spatial
structure of hadrons, one should notice that the transverse spatial dependence appearing in \cite{Bur2} 
is, however, necessarily buried 
in the correlators expressions defining TMDs, and it can only be probed explicitly 
through exclusive measurements.
The concrete possibility of testing the ideas on the role of transversity other than in semi-inclusive 
measurements, that are not directly sensitive to (transverse) spatial degrees of freedom, has been so far 
elusive. Suggestions were in fact made to obtain such information mainly from lattice 
calculations \cite{Bur2,Metz}.

The main thrust of this paper is on the contrary 
to propose a new avenue to experimentally determine transversity and its connection to spatial degrees of freedom, 
using exclusive processes such as $\pi^o$ and $\eta$ production off nucleons and nuclei. 
In this context two other functions, namely the combination of GPDs 
$2 \tilde{H}_T^q + E_T \equiv \overline{E}_T$, and the GPD $E$ defining the spin flip component in
scattering from an unpolarized nucleon, are essential. 
$\overline{E}_T$ is expected, in a class of models (\cite{Metz} and references therein), 
to be related to the Boer-Mulders function, $h_1^{\perp \, q}$ through: 
\begin{subequations}
\begin{eqnarray}
\label{E2}
\int\limits_{\zeta-1}^1 dX \, \left[ 2 \tilde{H}_T^q(X,\zeta,t) + E_T^q(X,\zeta,t) \right]_{\zeta=t=0}  = %
\kappa_T^q \\
\int d^2 k_T \, dX h_{1}^{\perp \, q}(X,k_T) \approx  - \kappa_T^q 
\end{eqnarray}
\end{subequations}
Similarly, $E$, can be related to the Sivers function, $f_{1T}^\perp$:
\begin{subequations}
\begin{eqnarray}
\label{E}
\int\limits_{\zeta-1}^1 dX \, \left[ E(X,\zeta,t) \right]_{\zeta=t=0} = \kappa^q \\
\int d^2 k_T \, dX f_{1T}^{\perp \, q}(X,k_T) \approx - \kappa^q  
\end{eqnarray}
\end{subequations}
Here, $\kappa^q$ is the contribution to the proton anomalous magnetic moment of the 
quark $q$. 
In the transverse plane in coordinate space, $\kappa^q$ is a measure of an unpolarized quark's 
displacement along the $y$-axis in a proton polarized along the $x$-axis \cite{Bur}.  
$\kappa_T^q$, the transverse anomalous moment, measures the $y$-transverse displacement for 
a transversely polarized quark along the $x$-axis in an unpolarized proton \cite{Bur2,BofPas}. 

The relationships above express at a deeper level the fact that both pairs: 
$[ (k_1+ik_2) f_{1T}^\perp(X,k_T), (\Delta_1+i \Delta_2)E(X,k_T;\zeta,\Delta) ]$ and 
$[(k_1+ik_2) h_{1}^\perp(X,k_T), (\Delta_1+i \Delta_2) \overline{E}_T(X,k_T;\zeta,\Delta) ]$, 
have the same helicity structure.
\footnote{$E(X,k_T;\zeta,\Delta)$ and $\overline{E}_T(X,k_T;\zeta,\Delta)$ are the unintegrated 
over $k_T$ GPDs.}
They can therefore be obtained from the overlap of similar wave functions at the handbag level.
Nevertheless, upon integration in $k_T$, $E$ does not necessarily vanish, while $f_{1T}^\perp$,
because of it is $T$-odd, is non zero only if FSI is taken into account.   
This is the reason why in the literature \cite{Bur2,Metz} no direct relationship was written
explicitly.

In order to make a connection with observables for given processes 
one needs to transform the Dirac matrix elements in terms 
of which the defining quark correlation functions 
are written, to the quark chirality basis \cite{JafJi,BogMul}. One can see that 
$H_T(X,\zeta,t)$ appears in the off-forward helicity amplitude $A_{++,--}$ \cite{Diehl_01} 
while $\overline{E}_T(X,\zeta,t)$ is defined in $A_{++,+-}, A_{-+,--}$ (more details are given below).
\footnote{We use the notation: $A_{\Lambda^\prime \lambda^\prime, \Lambda \lambda}$, where 
$\Lambda (\Lambda^\prime)$ are the initial (final) proton helicities, and  $\lambda (\lambda^\prime)$
are the initial (final) quark helicities}

As known from the helicity amplitude decomposition of observables in 
$\pi^0$ photoproduction~\cite{GolOwe},  
measurements of either the polarized target asymmetry, $A$, or the recoil nucleon polarization, 
$T$, or polarized photon 
asymmetries, $P$ on nucleons can be used to determine the 
desired helicity 
structure  
which we will see is connected to the off-forward quark helicity amplitudes (and related GPDs). 
A particular model was developed
in Ref.\cite{GolOwe} that, by accounting for the Regge cut corrections to the pole dominated amplitude, 
predicted
non-negligible values of $A$, $T$ and $P$ in the range $0 \lesssim -t \lesssim 1.$ GeV$^2$.  
The model contains the nucleons' tensor charges as parameters -- or a combination of parameters.
By extending this model to the case of virtual photon scattering
(Fig.\ref{fig1}b), one can, therefore, extract the value of the tensor charge 
directly from the data
using both the ideal intermediate energy kinematical range accessible at Compass, Hermes and Jefferson Lab, 
and the high precision provided by the latter.  
From the parton perspective, recent developments \cite{AHLT1,AHLT2} enable us to propose 
a detailed model where the quark 
degrees of freedom are described in terms of chiral-odd GPDs.

The connections being explored here, among transversity distributions, Regge exchange models and GPDs had 
some early hints from 
an exchange model that was developed for the tensor charge~\cite{GG1}. There the isovector and isoscalar 
components of the 
tensor charge were shown to be proportional to the product of the matrix elements for the decay of the 
lightest axial vector 
mesons ($J^{PC}=1^{+-}$), $b_1$ and $h_1$ respectively, and for the coupling of these mesons to the 
nucleon (Fig.\ref{fig1}b).
Both of them are needed in order to determine the isovector and isoscalar components of the tensor charge. 
Their coupling constants that entered the calculation were 
determined through the better known
decay constant of the $a_1$ ($J^{PC}=1^{++}$) meson, by 
exploiting the fact that $a_1$  belongs to the same $SU(6)\otimes O(3)$ multiplet as $b_1, h_1$ \cite{GG1}. 

It is important to notice that the $a_1$ ($J^{PC}=1^{++}$) type exchanges do not enter directly
the $\gamma^*p \rightarrow \pi^o p$ scattering amplitude. This will bear important consequences also 
in the description in terms of GPDs, namely in the identification of the correct structures entering the 
different helicity amplitudes.    
 

In the single meson or axial vector dominance approximation of Ref.\cite{GG1}, 
angular momentum conservation at point-like interaction 
vertices required that the coupling vanish in the forward limit. In order for the tensor charge not to vanish, a duality 
picture was envisaged where the struck quark undergoes Final State Interactions (FSI)  whose effect is parametrized 
in \cite{GG1}
in terms of the meson's constituent quarks' $\langle k_T^2 \rangle$.  
This interesting point of departure
from other treatments is the appearance of the factor $\left< k_T^2 \right>$ in the expression for the tensor
charge $\delta q$. This arises because of the kinematic
structure of the exchange picture that was adopted. 

The approach to the tensor charge just summarized leaves several questions - two in particular. Is the extrapolation 
from the $b_1$ mass ($t = m_{b_1}^2$) to $t=0$ a reasonable one? Can FSI's enter the picture in a natural way to 
parallel the $\left<k_T^2\right>$ dependence that was needed? Both of these questions are answered affirmatively 
in a Regge 
exchange model for $\pi^0$ electroproduction. The $b_1$ Regge trajectory enables the extrapolation 
to the $t \leq 0$ region. 
Regge cuts restore the exchange amplitude in the forward limit, as shown in photoproduction analyses in the 
past~\cite{GolOwe}. 
The GPD, or off-forward amplitude approach also provides answers to these questions. GPDs are functions of 
off-forward two body 
kinematic invariants, including $t$, and  allow for smooth extrapolation to $t=0$. Furthermore, the relevant 
off-forward helicity amplitude need not vanish in the forward limit, as will be seen later. 

A central question of factorization arises in relating the $\pi^0$ electroproduction single 
spin asymmetries to the relevant GPDs, because the important helicity amplitudes involve transverse virtual photons.
Factorization in meson production was explained in Ref.\cite{CFS} for longitudinally polarized photons only, 
based on the fact that in this case the end-point effects from the wave function of the produced meson 
are suppressed with respect to the transverse case \cite{BroFraGun}. 
This point of view was reiterated in several papers ~\cite{CF,Mank,AniTer,Pire,DieVin,HuaKro}.
On the other hand, GPD factorization was criticized recently by the authors of Ref.~\cite{SzcLlaLon} where it 
was claimed
that the dominance of Regge type exchanges produces non-analytic terms that destroy factorization. 
This is an ongoing and 
important debate now underway. Since our starting point is 
a ``duality'' type of picture, we think that there is theoretical merit 
in both perspectives. 

On the experimental front, however, there seems to be little evidence in the data that one 
can rule out transverse vs. longitudinal factorization
hypothesis. This appears in the $Q^2$ dependence of the 
recent HERMES $\rho$ production data showing a plateau in $R=\sigma_L/\sigma_T$ for vector meson
production \cite{Hermes_rho} (intermediate energy), \cite{Hera_rho} (high energy). 
We propose a new mechanism to describe the $Q^2$ dependence at the meson vertex that 
distinguishes the longitudinal and transverse photon polarization
contributions. 
This mechanism describes both the vector (natural parity) 
and axial vector  (unnatural parity)  channels 
taking into account  the orbital angular momentum
in the evaluation of the different quark helicity contributions to the pion wave function. 
In particular, the axial vector and vector channel differ by one unit of 
orbital angular momentum. Hence, as explained in detail later on,  the longitudinal photon amplitude is dominated 
by the axial vector contributions. For the transverse photon both vector and 
axial vector channels will contribute.

Our results, differing from Perturbative QCD (PQCD) type behavior, provide a less steep dependence on $Q^2$ of the longitudinal to transverse ratios.
We would like to remark that independent of the way corrections to the standard PQCD approach are carried out,  and of the interest in the process $per se$ 
as a probe of transversity, the issue of carefuly monitoring 
$\pi^0$ electroproduction at intermediate energies will be a 
prominent one in the analysis of many planned experiments at both Compass and Jlab. $\pi^0$'s constitute, in fact,  a large 
background in $\gamma$ production from both protons and 
nuclei.    

Finally, we suggest a practical method to extract both the tensor charge,  $\delta q$, and 
the transverse anomalous moment, $\kappa^q_T$,
from experiment that makes use of the fact that such quantities enter as free parameters in both our Regge
and partonic descriptions. We suggest a number of observables, {\it e.g.} the longitudinal/transverse 
interference term, $\sigma_{LT}$, the transverse spin asymmetry, $A_{UT}$, the beam spin polarization,
that are sensitive to the values of $\delta q$ and $\kappa^q_T$, $q=u,d$. 



The paper is organized as follows: in Section \ref{sec-2} we present definitions and kinematics; in Section 
 \ref{sec-3} we present our Regge approach;  
in Section \ref{sec-4} we introduce the connection to GPDs and we develop a parametrization of the 
chiral-odd ones, based on available experimental and phenomenological information; 
in Section \ref{sec-5} we propose a model of the $Q^2$ dependence of the various helicity amplitudes; 
in Section \ref{sec-6} we discuss results and propose an extraction method for the 
tensor charge; finally in Section \ref{sec-7} we draw our conclusions and present an outlook.
The spirit of the paper is to suggest a method to obtain $\delta q$ and  $\kappa_T^q$ from experiment, while
exploring a number of questions: from the dominance of chiral-odd contributions in $\pi^o$ electroproduction, 
to the duality picture, and the transition from Regge to partonic contributions. 

\section{$t$-channel dominance picture}
\label{sec-2}
In the following discussion of exclusive $\pi^0$ electroproduction, the crucial nexus connecting observable 
quantities, 
transversity GPDs and the Regge description is provided by helicity amplitudes. 
The relation between the electroproduction, virtual photon helicity amplitudes, 
$f_{\Lambda_\gamma,\Lambda_N; 0, \Lambda_N\prime}$  and the relevant GPDs is expressed in the ``handbag'' 
picture, 
with the assumption that the 
``hard part factorizes from the ``soft part'', as shown in Fig.~\ref{fig1}a. 
While this has been shown for the case 
of a high $Q^2$ longitudinal photon in exclusive vector meson electroproduction, it has not been demonstrated 
for transverse 
photons in $\pi^0$ production. The transverse photon contribution is kinematically suppressed 
by $1/Q$ relative to the 
longitudinal photon from the lepton tensor and the Hand convention for the virtual longitudinal photon flux. 
It has been presumed by many authors, without proof, that the transverse case does not factorize. 
Several papers have been written to 
carry the longitudinal case to higher twist, speculating that factorization still holds, while ignoring 
the transverse case, which should enter at the same order of $1/Q$ as twist 3 for longitudinal photons. 
Furthermore, the transverse virtual photon 
cross section does not show evidence of a decreasing ratio of transverse to longitudinal, 
as seen at HERMES~\cite{Hermes_rho} 
and in preliminary data from JLab~\cite{kubarovsky}, which leads us to reconsider the 
transverse case and to assume that 
factorization does hold.

We therefore start out by defining  the helicity amplitudes and the main observables related to transversity
within the assumption of factorization for the partonic description. 
It is important to recall that in the Regge pole or single particle exchange picture 
there is factorization of the upper 
vertex ($\gamma^*\rightarrow \pi^0$) from the lower vertex ($p\rightarrow p\prime$). 
This factorization would correspond 
to the $t$-channel picture (as in Fig.~\ref{fig1}) in which the quark and antiquark exchanges are accompanied by 
ladder-like gluon links. The Regge pole coupling at the vertex is independent of the other vertex and 
satisfies parity conservation.

\subsection{Kinematics and Definitions}
\label{subsec-2.1}
Exclusive $\pi^0$ electroproduction is shown in Fig.\ref{fig1}. The relevant four-momenta written in the 
laboratory frame are the initial
(final) electrons: $k_{1(2)} \equiv (\epsilon_1,{\bf k_{1(2)}})$, the exchanged photon's: $q \equiv (\nu,\bf{q})$,
with $\nu=\epsilon_1-\epsilon_2$, and ${\bf q} = {\bf k_{1}} -{\bf k_{2)}}$; the initial proton, 
$P\equiv (M,0))$. In addition, one has the final proton $P^\prime$, and the final pion $p_\pi$. 
We define the usual invariants for a Deeply Inelastic Scattering (DIS) process: the virtual photon's 
four-momentum squared, 
$Q^2= -q^2 = 4 \epsilon_1 \epsilon_2 \sin^2 \theta/2$, $(Pq)/M = \nu$, $x_{Bj} = Q^2/2M\nu$, $y=(Pq)/(P k_1)$. 
The Mandelstam invariants are defined with respect to the $\gamma^* N \rightarrow \pi^0 N^\prime $ process, namely:
$s = W^2 =(P+q)^2$, $t=(P-P^\prime)^2$, and $u=(P-p_\pi)^2$.    

The amplitudes are decomposed into a purely leptonic part and the  
$\gamma^*+ N \rightarrow \pi^0 + N^\prime $ process for which there are six 
independent helicity amplitudes chosen as 
\begin{equation}
f_1 = f_{1+,0+} \propto\Delta^1 ,\ \
f_2 = f_{1+,0-} \propto\Delta^0 ,\ \
f_3 = f_{1-,0+} \propto\Delta^2 ,\ \
f_4 = f_{1-,0-} \propto\Delta^1,
\label{f1}
\end{equation}
for transverse photons and
\begin{equation}
f_5 = f_{0+,0-} \propto\Delta^1, \ \
f_6 = f_{0+,0+} \propto\Delta^0.
\label{f0}
\end{equation}
for longitudinal photons. We use the notation: $f_{\Lambda_\gamma,\Lambda_N; 0, \Lambda_{N\prime}}$,
$\Lambda_\gamma = \pm 1,0$ being the virtual photon spin, and $\Lambda_N (\Lambda_{N\prime}) = +,- \equiv
+1/2,-1/2$ being the initial (final) nucleon spins; 
here $\Delta \equiv \mid {\bf \Delta} = {\bf P} - {\bf P}^\prime \mid$ is the magnitude of the three-momentum transfer, and the minimum kinematically
allowed power is indicated for each helicity amplitude. In a single hadron
exchange (or Regge pole exchange) factorization and parity conservation require 
\begin{equation}
f_1 =\pm f_4 \ {\rm and} \ f_2 =\mp f_3 
\end{equation}
for even or odd parity exchanges, to leading order in $s$. These pair relations, along with a
single hadron exchange model, force $f_2$ to behave like $f_3$ for small
$\Delta$ or small transverse momentum for the outgoing particles. This introduces the $k_T^2$ factor into the $f_2$ 
amplitude, which is related to the transversity transfer, as we will see below from its connection with the GPD $H_T$. 
Observable quantities are bilinear combinations of these helicity amplitudes. 

The differential cross section for pion electroproduction off an unpolarized target is \cite{RasDon}
\begin{equation} 
\label{unp-xs}
\frac{d^4\sigma}{d\Omega d\epsilon_2 d\phi dt} = \Gamma \left\{ \frac{d\sigma_T}{dt} + \epsilon_L \frac{d\sigma_L}{dt} + \epsilon \cos 2\phi \frac{d\sigma_{TT}}{dt} 
+ \sqrt{2\epsilon_L(\epsilon+1)} \cos \phi \frac{d\sigma_{LT}}{dt} \right\}.
\end{equation}
If the initial electron is polarized, with $h=\pm 1$, one has the additional contribution
\begin{equation}
\label{pol-xs} 
h  \, \sqrt{2\epsilon_L(\epsilon-1)} \, \frac{d\sigma_{L^\prime T}}{dt} \sin \phi ,
\end{equation}
The photon polarization parameter $\epsilon$ can be written in terms of invariants as

\[ \epsilon^{-1} = 1 + 2\left( 1+\frac{\nu^2}{Q^2} \right)\left(4 \dfrac{\nu^2}{Q^2} \dfrac{1-y}{y^2}-1\right)^{-1}, \] 

\noindent and for longitudinal polarization alone, 

\[ \epsilon_L=  \frac{Q^2}{\nu^2} \epsilon. \]

\noindent 
The factor $\Gamma$ is given by
\begin{equation}
\label{Gamma}
\Gamma = \sigma_{Mott} \, f_{rec} \, m_\pi \mid {\bf p}_\pi \mid J(Q^2,\nu,s) 
\end{equation}

\noindent 
where the Mott cross section is
\[ \sigma_{Mott}= \frac{4 \alpha^2 \epsilon_2^2 \cos^2 \theta/2}{Q^4},  \]  

\noindent
and the hadronic recoil factor, $f_{rec}$,

\[ f_{rec} = \lvert 1 + \frac{\nu - \mid {\bf q} \mid \cos \theta_\pi^{LAB}}{M} \rvert^{-1}. \]

\noindent 
In Eq.(\ref{Gamma}), $J(Q^2,\nu,s)$ is the jacobian for the transformation from $\cos \theta_\pi^{LAB}$ to
$t$, whose expression is given in  Appendix \ref{app-1}.

The different contributions in Eq.(\ref{unp-xs}) are written in terms of helicity amplitudes as
\begin{eqnarray}
\label{dsigT}
\frac{d\sigma_T}{dt} & = & \mathcal{N} \, %
\left( \mid f_{1,+;0,+} \mid^2 + \mid f_{1,+;0,-} \mid^2 + \mid f_{1,-;0,+} \mid^2 +
\mid f_{1,-;0,-} \mid^2 \right)  \nonumber \\
& = & \mathcal{N} \, \left( \mid f_1 \mid^2 + \mid f_2 \mid^2 + \mid f_3 \mid^2 +
\mid f_4 \mid^2 \right)  
\\
\label{dsigL}
\frac{d\sigma_L}{dt} & = & \mathcal{N}  \left(\mid f_{0,+;0,+} \mid^2 + \mid f_{0,+;0,-} \mid^2 \right) \nonumber \\
& = & \mathcal{N} \,  \left( \mid f_5 \mid^2 + \mid f_6 \mid^2 \right), 
\end{eqnarray}

for transverse and longitudinal virtual photon polarizations, respectively, and
\begin{equation}
\label{Hand}
 \mathcal{N} = \left[ M (s-M^2)^2 \right]^{-1} \, G 
\end{equation}
where we used the Hand convention, multiplied by a geometrical factor $G$ that is given by 
$G=\pi/2$, and $G=1/8\pi$, in the Regge and GPD approaches considered later on.

The cross section for the virtual photon linearly polarized out of the scattering plane minus 
that for the scattering plane is 
\begin{eqnarray}
\label{dsigTT}
\frac{d\sigma_{TT}}{dt} & = & 2 \, \mathcal{N}  \Re e \left( f_{1,+;0,+}^*f_{1,-;0,-} - f_{1,+;0,-}^* f_{1,-;0,+} \right)
\nonumber \\
& = & 2 \, \mathcal{N} \,    \Re e \left( f_1^*f_4 - f_2^* f_3 \right).
\end{eqnarray}

The interference term for the transversely and longitudinally polarized virtual photons is 
\begin{eqnarray}
\frac{d\sigma_{LT}}{dt} & = & 2 \, \mathcal{N} \,  
\Re e \left[ f_{0,+;0,+}^* (f_{1,+;0,-} + f_{1,-;0,+}) + f_{0,+;0,-}^* (f_{1,+;0,+} - f_{1,-;0,-}) \right]
\nonumber \\
& = & 2 \, \mathcal{N} \, 
\Re e \left[ f_5^* (f_2 + f_3) + f_6^* (f_1 - f_4) \right].
\label{dsigLT}
\end{eqnarray}
Finally the beam polarization term is given by
\begin{eqnarray}
\frac{d\sigma_{LT^\prime}}{dt} & = & 2 \, \mathcal{N} \,  
\Im m \left[ f_{0,+;0,+}^* (f_{1,+;0,-} + f_{1,-;0,+}) + f_{0,+;0,-}^* (f_{1,+;0,+} - f_{1,-;0,-}) \right]
\nonumber \\
& = & 2 \, \mathcal{N}  \, 
\Im m  \left[ f_5^* (f_2 + f_3)  + f_6^* (f_1 - f_4) \right]
\label{dsigLTp}
\end{eqnarray}

In addition to the unpolarized observables listed above, a number of observables
directly connected to transversity can be written (see {\it e.g.} \cite{GolMor}). 
Here we give the transversely polarized target asymmetry,
\begin{equation}
\label{AUT}
A_{UT} =  \frac{2 \Im m (f_1^*f_3 - f_4^*f_2)}{{\displaystyle \frac{d\sigma_T}{dt}}},
\end{equation}
and the beam spin asymmetry,
\begin{equation}
\label{BSA}
A \approx  \alpha \sin \phi,
\end{equation}
where 
\begin{equation}
\label{alpha}
\alpha = \frac{\sqrt{2 \epsilon_L (1-\epsilon)} \,\, {\displaystyle \frac{d \sigma_{LT^\prime}}{dt} }}%
{{\displaystyle \frac{d\sigma_T}{dt} + \epsilon_L \frac{d\sigma_L}{dt} } }
\end{equation}
(note that the recoil nucleon polarization asymmetry, $\mathcal{T}$, 
defined analogously to $A_{UT}$ simply involves the switching of $f_1$ and $f_4$).

It is important to realize that the relations between observables and helicity amplitudes 
are general, independent of any particular model. We will see that the Regge model, as well 
as the parameterization through GPDs, populate those helicity amplitudes related to transversity 
and thereby effect observables in important ways. 

\subsection{Connection to Generalized Parton Distributions}
\label{subsec-2.2}
The connection with the parton model and the transversity distribution is uncovered through the GPD 
decomposition of the helicity amplitudes.
In the factorization scenario the amplitudes for exclusive $\pi^0$ electroproduction, 
$f_{\Lambda_\gamma,\Lambda_N; 0, \Lambda_{N\prime}}$ 
can be decomposed into a ``hard part'', 
$g_{\Lambda_\gamma,\lambda; 0, \lambda^\prime}$ describing the partonic subprocess 
$\gamma^* + q \rightarrow \pi^0 + q$ (top part of diagram in Fig\ref{fig1}a), 
and a ``soft part'', $A_{\Lambda^\prime,\lambda^\prime;\Lambda,\lambda}$ 
that, in turn, contains the GPDs,
\begin{equation}
f_{\Lambda_\gamma,\Lambda;0,\Lambda^\prime} = \sum_{\lambda,\lambda^\prime} 
g_{\Lambda_\gamma,\lambda;0,\lambda^\prime} (X,\zeta,t,Q^2)  \otimes
A_{\Lambda^\prime,\lambda^\prime;\Lambda,\lambda}(X,\zeta,t),
\label{facto}
\end{equation}
where a sum over the different quark components is omitted for simplicity. 
The amplitudes in  Eq.(\ref{facto}) implicitly contain an integration over the unobserved quark momenta, 
and are functions of 
$x_{Bj} \approx \zeta, t$ and $Q^2$; they are analogous to the Compton Form Factors in DVCS.
In fact, in the more familiar case of DVCS, the upper part involves the matrix elements of $j_\mu^{EM}(x) j_\nu^{EM}(0)$, the tree level diagram for $\gamma^*+q \rightarrow \gamma + q$ with the high momentum struck quark intermediate state. In the Bjorken limit, in terms of light cone variables, this diagram has a simple Dirac structure of $\gamma^+$ with a denominator that becomes $1/(X-i\epsilon)$ (the corresponding crossed diagram, required for gauge invariance, yields $1/(X -\zeta+i\epsilon)$). 
The same structure would obtain for the production of vector mesons -- matrix elements of $j_\mu^{EM}(x) j_\nu^V(0)$), along with a hadronic wavefunction.
One model for this, that is often used, has the hadronic wave function nicely factored from the partonic components of the PQCD diagrams. With this latter approach there is a  depression of longitudinal to/from transverse transitions~\cite{CFS,Pire}. For pion production the upper, hard part of the diagram involves the matrix element between quark states of $j_\mu^{EM}(x) j^P(0)$, the latter being the pseudoscalar hadronic current operator. As we have emphasized, for $\pi^0$ the diagram is C-Parity odd and chiral odd in the $t-$channel. 
Because only one of these transverse photon functions survives the limits, the relation to  
the tensor charge is 
quite simple. Note that because of the pion chirality (0$^-$), the quark must flip helicity at the pion vertex where we take the coupling 
to be $\gamma^5$. 
Therefore, the corresponding Dirac structure for the hard subprocess diagram involves $\sigma^{+T}\gamma^5$, 
at variance with Refs.\cite{Mank,VGG} where the C-parity even axial vector structure
$\gamma^\mu\gamma^5$ was considered. This is very significant for our reaction.
Our observation also implies important changes in the $Q^2$ dependence of the process that will be discussed within a specific model in Section V.  

As we displayed for the t-channel picture of electroproduction, there are six independent helicity amplitudes 
for $\gamma^*+N\rightarrow\pi^0+N^\prime$, given parity conservation, four with $\lambda_\gamma=1$, Eq.~(\ref{f1}) 
and two with $\lambda_\gamma=0$, Eq.~(\ref{f0}). It will be important in the following to observe that each helicity 
amplitude $f_{\Lambda_\gamma,\Lambda;0,\Lambda^\prime}$ 
will have an angular momentum conserving factor of $sin^n(\theta_{CM}/2)$, where the minimum value of 
$n=\Lambda_\gamma-\Lambda+\Lambda^\prime$. This was written in terms of powers of $\mid {\bf \Delta} \mid $ $\propto sin(\theta_{CM}/2)$.
in Eqs.~(\ref{f1}) and ~(\ref{f0}). 
In terms of the invariant variables used here for the GPDs, 
\begin{equation}
sin^2(\theta_{CM}/2) = -\frac{\zeta}{Q^2} (t - t_{min})
\end{equation}
where the limit of $Q^2 >> M^2$ is taken. Corresponding factors of $sin^{n^\prime}(\theta_{CM}/2)$ will 
occur in the $g$ and $A$ amplitudes.

The helicity structure of the $g$ amplitudes is straightforward. It is the same as 
the $f$ amplitudes, so we can take the same labeling for $g_1,...,g_6$ as in Eqs.(\ref{f1}) and (\ref{f0}). Using parity conservation,
$g_1= g_{1+,0+}$,  $g_2= g_{1+,0-}$, $g_3= g_{-1+,0-} = g_{1-,0+}$, $g_4= g_{1-,0-}$, $g_5= g_{0+,0-}$,  and $g_6=g_{0+,0+}$. 
If we have 
$${\rm \gamma^*(q) + q(k)\rightarrow \pi^0(p_\pi) + q(k^\prime)},$$ 
then $\hat{s}=(q+k)^2, t=(k^\prime-k)^2 = (q-p_\pi)^2, \mid {\bf \Delta} \mid^2=t_{min}-t$, with $t_{min}$ fixed in the 
center of mass (CM) of this reaction, and $\hat{u}=(q-k^\prime)^2$. 
With these variables, and taking the quarks and pion masses to zero, the $g$ amplitudes are quite simple. Only $g_2$ and $g_5$ survive in the $\hat{s}>>|t|$ limit.
One has,
\begin{subequations}
\label{g_functions}
\begin{eqnarray}
g_1 & = &  g_4=  g_\pi \, \frac{\mathcal{C}_q }{\hat{s}}  
\frac{1}{4}  N \, N^\prime \,  {\rm Tr}  \left \{\gamma^\lambda \gamma^o(1 + \gamma^3\gamma^5) \gamma^\mu \gamma^+ \gamma^\nu \right\} k_\lambda k^\prime_\mu \epsilon_\nu^T  \nonumber 
\\
& = &  g_\pi \, \frac{\mathcal{C}_q }{\hat{s}}
N \, N^\prime \left[ \epsilon^{\lambda 3 \mu \nu} +   \epsilon^{\lambda o \mu \nu} + (g^{\mu o} - g^{\mu 3} ) \epsilon^{\lambda o 3 \nu}  \right] k_\lambda k^\prime_\mu \epsilon_\nu^T = 0  \\
g_2 & = &   g_\pi \, \frac{\mathcal{C}_q }{\hat{s}}  
\frac{1}{4}  N \, N^\prime \,  {\rm Tr}  \left \{\gamma^\lambda \gamma^o(-\gamma^1 + i\gamma^2)\gamma^\mu \gamma^+ \gamma^\nu \right\} k_\lambda k^\prime_\mu \epsilon_\nu^{(\lambda=+1)}  \nonumber 
\\
& = &  g_\pi \, \mathcal{C}_q  \cos \theta/2 \simeq g_\pi \, \mathcal{C}_q  \sqrt{\frac{-\hat{u}}{\hat{s}}} \\
g_3 & = & g_\pi \, \frac{\mathcal{C}_q }{\hat{s}}  
\frac{1}{4}  N \, N^\prime \,  {\rm Tr}  \left \{\gamma^\lambda \gamma^o(-\gamma^1 + i\gamma^2)\gamma^\mu \gamma^+ \gamma^\nu \right\} k_\lambda k^\prime_\mu \epsilon_\nu^{(\lambda=-1)}  =0  \\
g_5 & = &   g_\pi \, \frac{\mathcal{C}_q }{\hat{s}}  
\frac{1}{4}  N \, N^\prime \,  {\rm Tr}  \left \{\gamma^\lambda \gamma^o(-\gamma^1 + i\gamma^2)\gamma^\mu \gamma^+ \gamma^\nu \right\} k_\lambda k^\prime_\mu \epsilon_\nu^L  \nonumber  \\
& = &  g_\pi \, \mathcal{C}_q  \frac{\sqrt{2\hat{s}}}{Q} \sin \theta/2\simeq g_\pi \, \mathcal{C}_q  \sqrt{\frac{-2t}{Q^2}}, \\
g_6 & = & g_\pi \, \frac{\mathcal{C}_q }{\hat{s}} 
\frac{1}{4}  N \, N^\prime \,  {\rm Tr}  \left \{\gamma^\lambda \gamma^o(1 + \gamma^3\gamma^5) \gamma^\mu \gamma^+ \gamma^\nu \right\} k_\lambda k^\prime_\mu \epsilon_\nu^L = 0
\end{eqnarray}
\end{subequations}
where the photon polarization vectors are 
$\epsilon_\nu^T = \epsilon_\nu^{(\lambda=\pm 1)} \equiv 1/\sqrt{2} (0;\mp 1,- i,0)$,  $\epsilon_\nu^L=\epsilon_\nu^{(\lambda=0)} \equiv 1/\sqrt{Q^2} (\mid {\bf q} \mid; 0_\perp,\nu)$, and 

\[  \mathcal{C}_q  = \frac{1}{X- i \epsilon} + \frac{1}{X-\zeta + i \epsilon}  \]

\noindent 
$g_\pi \simeq \sqrt 15 (2/3)$ is the quark-pion 
coupling obtained from the nucleon-pion coupling in the additive quark model: we take this limiting value to show
the structure of  the upper part of the handbag. 
The composite structure of the pion production vertex generates an additional $Q^2$ dependence that will be described in
detail in Section V.  
Our goal is to provide an alternative to the standard PQCD based meson production models that are well known to
largely miss the behavior of current experiments in the Multi-GeV kinematical region (see however discussion in \cite{GolKro}).   
We obtain different $Q^2$ behaviors for the subprocess amplitudes, corresponding to either
axial vector (A), or vector (V) $t$-channel exchanges, or depending, in other words, on the $C$ and $P$ quantum numbers.
In our model this translates into a different dependence for each one of the helicity amplitudes entering Eq.(\ref{facto}),
\begin{subequations}
\label{f15}
\begin{eqnarray}
f_1  =  f_4 & = & \int \limits_{-1+\zeta}^{1} dX \, g_2 (X,\zeta,t,Q^2)  \, F_V(Q^2) \, A_{++,+-}(X,\zeta,t) \\
f_2 & =  & \int \limits_{-1+\zeta}^{1} dX \, g_2 (X,\zeta,t,Q^2)  \, [F_V(Q^2)+F_A(Q^2)] \, A_{--,++}(X,\zeta,t) \\
f_3 & = & \int \limits_{-1+\zeta}^{1} dX \, g_2 (X,\zeta,t,Q^2)  \, [F_A(Q^2)-F_V(Q^2)]  \, A_{+-,-+}(X,\zeta,t) \\
f_5 & = & \int \limits_{-1+\zeta}^{1} dX \, g_5 (X,\zeta,t,Q^2)  \, F_A(Q^2)  \, A_{--,++}(X,\zeta,t) 
\end{eqnarray}
\end{subequations}
Note that $A_{--, ++}$ involves no overall helicity change, and hence no required factor 
of a non-zero power of $ t - t_0$. Nevertheless, many models will lead to non-zero powers 
that kill the forward limit, and hence, do not contribute to the tensor charge.
This point is explained in detail below for the Regge model adopted in this paper.


\section{Weak-Cut Regge Model}
\label{sec-3}
To have a non-zero single spin asymmetry requires interference between
single helicity flip and non-flip and/or double flip amplitudes. Asymmetry
arises from rescattering corrections (or Regge cuts or eikonalization or
loop corrections) to single hadron exchanges. That is, one of the amplitudes in the product must acquire a 
different phase, a relative imaginary part. We will construct a Regge pole model with cuts to account for 
measured photoproduction 
observables at moderate s and small t, and extend the model into electroproduction for moderate $Q^2$.

We now employ a Regge pole description of the $\pi^0$ electroproduction, following the ``weak cut'' approach used by 
Goldstein and Owens \cite{GolOwe} where all the observable quantities are defined in terms of the $f$ amplitudes, as related above. 
From \cite{GolOwe} one can immediately see that the specific combination that contains $H_T$ 
is the one given above because this implies axial-vector t-channel exchanges. 

The leading Regge trajectories that are exchanged in this or any two-body diffractive process 
can be categorized by the signature and parity. The signature determines whether the poles in the exchange 
amplitudes will occur 
for even or odd positive integer values of the spin trajectory $\alpha(t) = J$. 
The leading axial vectors are $b_1$ and $h_1$. These are crucial for determining 
the tensor charge and 
the transversity distribution.
The parameterization of these trajectories for each helicity amplitude takes the form of product of (1) a 
t-dependent coupling for both vertices, or the residue function, (2) a signature factor, dependent on the Regge trajectory, 
that determines the positions of the poles in the t-channel, and (3) the energy dependence, a power fixed by the trajectory. 
For the even signature exchanges, the natural parity $\rho$ and $\omega$, the amplitudes are
\begin{eqnarray}
f_1=f_4=\frac{\beta_1^V}{\Gamma(\alpha_V(t))} \Delta \frac{1-e^{-i\pi\alpha_V(t)}}{sin(\pi\alpha_V(t))}\nu^{\alpha_V(t)}e^{-\Delta^2c_V/2} \\
f_2=-f_3=\frac{-\beta_2^V \Delta^2}{2 M\Gamma(\alpha_V(t))} \frac{1-e^{-i\pi\alpha_V(t)}}{sin(\pi\alpha_V(t))}\nu^{\alpha_V(t)}e^{-\Delta^2c_V/2}
\end{eqnarray}
For the odd signature, unnatural parity $b_1$ and $h_1$, $f_1=f_4=0$,
\begin{eqnarray}
f_2=+f_3=\frac{\beta_1^A \Delta^2}{2 M\Gamma(\alpha_A(t)+1)} \frac{1-e^{-i\pi\alpha_A(t)}}{sin(\pi\alpha_A(t))}\nu^{\alpha_A(t)}e^{-\Delta^2c_A/2}.
\end{eqnarray}
The additional contributions to the longitudinal photon amplitudes are limited by parity, charge conjugation 
invariance and 
helicity conservation to  the $b_1$ and $h_1$ only. To leading order in $s$ they contribute only to $f_5$, which has the 
minimal $\Delta^1$ dependence and is proportional to the same trajectories' contributions to $f_2$. The precise relation 
between these two amplitude contributions depends on $Q^2$ and can be related to the decay widths for the axial vectors 
using vector dominance of the photon. The vector meson trajectories do not couple to the helicity zero photon and pion 
vertex.
Clearly all of these amplitudes vanish in the forward direction, as a result of the factorization of the Regge pole 
exchanges 
into two vertices and the fact that parity conservation holds separately for both vertices. This implies that $f_2$, 
which, 
in general has the minimal angular dependence of $\Delta^0$, for the Regge pole contributions alone, acquires the same 
angular dependence as $f_3$. Since the $\Delta\rightarrow 0$ limit is quite important for the identification with 
transversity through 
$A_{\Lambda,\lambda;\Lambda^\prime,\lambda^\prime}$ and thus through the GPD $H_T$, it behooves us to consider the 
rescattering or FSI in this picture. This leads to the Regge cut scheme for rescattering corrections, which was quite 
important for making contact with spin dependent data.

The Regge cut scheme is implemented by first taking the impact parameter representation of the pole terms as an eikonal, 
\begin{equation}
\chi_{Regge}(s,b)=\frac{1}{k\sqrt{s}}\int \Delta d\Delta J_n(b,\Delta)f(s,\Delta^2),
\label{eikonal}
\end{equation}
with $n$ being the helicity change. Next convoluting with the eikonal obtained from helicity conserving Pomeron 
exchange, 
$\chi_P(s,b)$ and then transforming back to the momentum space representation
\begin{equation}
f(s,\Delta^2)=ik\sqrt{s}\int b db J_n(b,\Delta)\chi_{Regge}(s,b)\chi_P(s,b).
\label{convolution}
\end{equation}
This restores the $\Delta^0$ behavior of $f_2$, thereby providing a non-zero tensor charge through the Regge couplings.

The tensor charge is embedded in the residues of these $1^{+-}$ axial vector Regge trajectories, $A$ generically. 
The residue for $b_1$ or $h_1$ is expressed in terms of coupling constants and other dynamical factors that arise from 
evaluating the residue at the pole position, $t=m_{A}^2$. There 
\begin{equation}
\beta_1^A=\frac{g_{\gamma A \pi} g_{AN\bar{N}}}{4\pi m_A}\frac{\pi \alpha_A^\prime}{4\sqrt{2}}e^{-m_A^2c_A/2},
\label{couplings}
\end{equation}
and the critical coupling is the $g_AN\bar{N}$ with factors from the Regge parameterization.

\section{Generalized Parton Distributions}
\label{sec-4}
In order to explore the connection between the previous formalism with a partonic 
picture one needs to rewrite the various observables 
listed in Eqs.(\ref{dsigT},\ref{dsigL},\ref{dsigTT},\ref{dsigLT},\ref{AUT})
in terms of the correlation
functions for the handbag diagram in Fig.\ref{fig1}a. 

The electroproduction amplitude for hard exclusive pseudoscalar meson production 
can be written in a factorized form where the soft, process independent, 
part involves both the description of the meson vertex
and linear combinations of GPDs at the nucleon vertex, as 
in the expression presented in Eq.(\ref{facto}).
A formal proof of factorization was given in the case of longitudinally polarized 
virtual photons producing longitudinally polarized vector mesons
\cite{CFS}.   
The proof 
hinges on the hypothesis that the initial quark-antiquark pair
produced in the hard interaction is in a pointlike configuration, thus granting the cancellation
of soft gluons contributions, and only subsequently evolving into the observed meson.
Endpoint contributions are surmised to be larger in electroproduction of 
transversely polarized vector mesons, and to therefore prevent factorization.     
Standard pQCD calculations \cite{BroFraGun} predict a ratio of $\sigma_L/\sigma_T \propto Q^2$. 
In \cite{Martinetal} it was observed that when this ratio is calculated in non-perturbative models, 
an even larger relative suppression of $\sigma_T$ might arise. 
Notwithstanding current theoretical approaches, many measurements 
conducted through the years, 
display larger transverse contributions than expected \cite{Hermes_rho,Hera_rho}.
In order to explain the $Q^2$ dependence of the large $W^2$ data in \cite{Martinetal} the hypothesis
of duality was used, whereby factorization was assumed, but the meson distribution amplitude 
was omitted.
On the other side, with analogous arguments as for other reactions measured in 
the multi-GeV region, limitations to the factorization scenario challenging the 
``point-like nature'' of the produced $q \overline{q}$ pair even in longitudinal
polarization scattering, were suggested in \cite{HoyLen}. 

\begin{figure}
\includegraphics[width=12.cm]{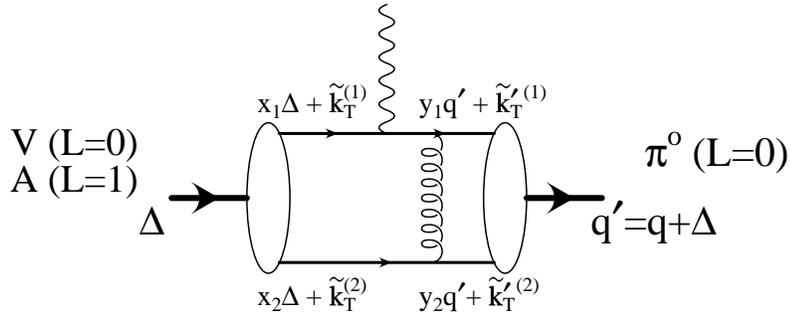}
\caption{Perturbative QCD contribution to the $\gamma^* V \pi^o$ and $\gamma^* A \pi^o$ 
form factors.} 
\label{fig2} 
\end{figure}

In the case of pseudoscalar ($\pi$ and $\eta$) production, 
preliminary data seem also to indicate transverse contributions 
larger than predicted within pQCD \cite{kubarovsky}.
Lacking a complete formal proof of factorization (see however \cite{Ivanov}), 
it is therefore important to explore alternative avenues 
for the meson production mechanism.
In this paper we suggest a QCD based model, described in more detail in Section \ref{sec-5}, 
that predicts different $Q^2$ behaviors for meson production via natural and unnatural
parity channels.These are defined in the upper part
of the diagram in Fig.\ref{fig2}, as $F_V(Q^2)$, and $F_A(Q^2)$, for the natural and unnatural
parity exchanges, respectively. In what follows we give the expressions for the various
terms in the $\pi^0$ electroproduction cross section in terms of GPDs and of the 
$Q^2$-dependent factors. The explicit form of the factors is explained in Section \ref{sec-5}.

\subsection{Kinematics and Definitions}
\label{sec-3_kin}
  
The off-forward correlation matrix is defined in the collinear approximation as
\footnote{Notice that we adopt the
axial gauge, although results can be cast in a form 
highlighting gauge invariance \cite{Ji1}.}  
\begin{eqnarray}
\label{corr1}
\Phi_{ab} = \int  \, 
\frac{ d y^-}{2 \pi} \,  
e^{i y^- X} \, \langle P^\prime S^\prime \mid \overline{\psi}_b(0)  \psi_a(y^-) \mid P S \rangle
\end{eqnarray}
where $P,S$ ($P^\prime,S^\prime$) are the initial (final) nucleon momentum and spin, and 
we wrote explicitly the Dirac indices $a,b$.
The C-parity odd and chiral odd quark density matrix we are interested in~\cite{RalSop,JafJi} involves the struck quark helicity flip via the contraction of $\Phi_{ab}$ with the Dirac matrix $(i \sigma^{+ \, i})_{ba}$.

\noindent This contraction, as shown in Ref.\cite{Diehl_01}, gives rise to four chiral odd GPDs, 
\begin{eqnarray}
\int dk^- \, d^2{\bf k} \,  
 \, \rm{Tr} \left[ i\sigma^{+i} \Phi \right]_{X P^+=k^+} & & \nonumber \\
  =   \frac{1}{2 P^+}  \, \overline{U}(P^\prime, S^\prime) \,
 [ H_T^q \, i \sigma^{+ \, i} + & \widetilde{H}_T^q  \, \frac{P^+ \Delta^i - \Delta^+ P^i}{M^2} & \, 
 +E_T^q \, \frac{\gamma^+ \Delta^i - \Delta^+ \gamma^i}{2M} \,
 \nonumber \\
 &   + \widetilde{E}_T^q \, \frac{\gamma^+P^i - P^+ \gamma^i}{M}]  & U(P,S) 
\label{oddgpd} 
\end{eqnarray}
where 
$q=u,d,s$. The nucleon spinors can be explicitly chosen in various ways. Diehl in Ref.~\cite{Diehl_01} chooses light-front helicity spinors to obtain 4 independent chiral odd amplitudes, the $A_{\Lambda^\prime,\lambda^\prime;\Lambda,\lambda}$'s in Eqn.~\ref{facto} and~\ref{f15}. These are linear combinations of the same 4 GPDs that appear in Eqn.~\ref{oddgpd}, namely
\begin{subequations}
\label{helamps_gpd}
\begin{eqnarray} 
A_{+-,++} & = & - \frac{\sqrt{t_0-t}}{2M} \left[ \widetilde{H}_ T + \frac{1+\xi}{2}E_T - 
\frac{1+\xi}{2}\widetilde{E}_T \right] 
\\
A_{++,--} & = & \sqrt{1-\xi^2} \left[ H_ T + \frac{t_0-t}{4M^2} \widetilde{H}_T%
- \frac{\xi^2}{1-\xi^2}E_T + 
\frac{\xi}{1-\xi^2}\widetilde{E}_T \right] 
\\
A_{+-,-+} & = & - \sqrt{1-\xi^2} \, \, \frac{t_0-t}{4M^2} \, \widetilde{H}_T   
\\
A_{++,+-} & = &  \frac{\sqrt{t_0-t}}{2M}  \left[ \widetilde{H}_T + \frac{1-\xi}{2}E_T 
+ \frac{1-\xi}{2}\widetilde{E}_T \right],
\end{eqnarray}
\end{subequations}
where for consistency with previous literature we have used $\xi=\zeta/(2-\zeta)$; $t_0 = - M^2 \zeta^2/(1-\zeta)$,
$M$ being the proton mass. 

The exclusive process observables are defined in terms of the helicity amplitudes of Eqn.~\ref{facto}, which involves the integration over $X$ of the $A$'s with the $g$'s. In the Bjorken limit $g_2$ and $g_5$ are independent of $X$, except for the propagator denominators contained in the $\mathcal{C}_q$. Hence the overall helicity amplitudes $f_1$ to $f_6$ involve the analog of the Compton Form Factors,  the Meson Production Form Factors (MPFFs) which can be written generically as~\cite{Die_rev,BelRad}:
\begin{eqnarray}
\mathcal{F}^q(\zeta,t)  = i \pi 
\left[ F^q(\zeta,\zeta,t) - F^{\bar{q}}(\zeta,\zeta,t) \right] +
\mathcal{P} \int\limits_{-1+\zeta}^1 dX \left(\frac{1}{X-\zeta} + \frac{1}{X} \right) F^q(X,\zeta,t).
\label{cal_F}
\end{eqnarray}
where $\mathcal{P}$ indicates a principal value integration and 
$\mathcal{F}^q = \mathcal{H}_T^q, \mathcal{E}_T^q, \widetilde{\mathcal{H}}_T^q, \widetilde{\mathcal{E}}_T^q$.
Notice that the $\Delta_T \equiv \sqrt{t_0-t}$ dependence in Eqs.(\ref{helamps_gpd}) is the same as the minimal
$\Delta_T$ dependence predicted within the Regge model (see \cite{GolOwe} and Section \ref{sec-2}).  
The MPFFs appearing in Eq.(\ref{helamps_gpd}) are:
\begin{equation}
\mathcal{F} \equiv \mathcal{F}^{p\rightarrow \pi^o} = \frac{1}{\sqrt{2}} \left[ \frac{2}{3} \mathcal{F}^u + 
\frac{1}{3} \mathcal{F}^d \right].
\end{equation}

To form the complete helicity amplitudes the $A$'s of Eqn.~\ref{helamps_gpd} are inserted into integrals over $X$ with the same form as Eqn.~\ref{cal_F},
\begin{eqnarray}
\mathcal{A}^q(\zeta,t)  = i \pi 
\left[ A^q(\zeta,\zeta,t) - A^{\bar{q}}(\zeta,\zeta,t) \right] +
\mathcal{P} \int\limits_{-1+\zeta}^1 dX \left(\frac{1}{X-\zeta} + \frac{1}{X} \right) A^q(X,\zeta,t).
\label{cal_A}
\end{eqnarray}
Being linear in the GPDs, the relations in Eqn.~\ref{helamps_gpd} are preserved for the integrated $A$'s and $\mathcal{F}$'s. This is due to the fact that the hard process factors, the $g_2$ and $g_5$, have only the $X$ dependence of their propagators, $\mathcal{C}_q$, which provide the denominators in the above integration over $X$. The complete helicity amplitudes, $f_1$ to $f_6$ are then given by the $\mathcal{A}$ multiplied by the $g_2/\mathcal{C}_q$ or $g_5/\mathcal{C}_q$ and the corresponding $Q^2$ dependent factors that are shown in Eqn.~\ref{f15}. The $Q^2$ dependent factors depend on the quantum numbers in the $t$-channel, and will be discussed in Section~\ref{sec-5}.

The observables defined in Section \ref{sec-2} can be written in terms of the $p \rightarrow \pi^0$ MPFFs  through the helicity amplitudes $f_1$ to $f_6$. 
\begin{subequations}
\label{helamps_gpd2}
\begin{eqnarray} 
f_1 = f_4 & = & \frac{g_2}{\mathcal{C}_q}F_V(Q^2) \frac{\sqrt{t_0-t}}{2M} \left[ \widetilde{\mathcal{H}}_ T + \frac{1-\xi}{2}\mathcal{E}_T + 
\frac{1-\xi}{2}\widetilde{\mathcal{E}}_T \right] 
\\
f_2 & = & \frac{g_2}{\mathcal{C}_q}[F_V(Q^2)+F_A(Q^2)]\sqrt{1-\xi^2} \left[ \mathcal{H}_ T + \frac{t_0-t}{4M^2} \widetilde{\mathcal{H}}_T%
- \frac{\xi^2}{1-\xi^2}\mathcal{E}_T + 
\frac{\xi}{1-\xi^2}\widetilde{\mathcal{E}}_T \right] 
\\
f_3 & = & \frac{g_2}{\mathcal{C}_q}[F_V(Q^2)-F_A(Q^2)] \sqrt{1-\xi^2} \, \, \frac{t_0-t}{4M^2} \, \widetilde{\mathcal{H}}_T   
\\
f_5  & = & \frac{g_5}{\mathcal{C}_q}F_A(Q^2) \sqrt{1-\xi^2} \left[ \mathcal{H}_ T + \frac{t_0-t}{4M^2} \widetilde{\mathcal{H}}_T%
- \frac{\xi^2}{1-\xi^2}\mathcal{E}_T + 
\frac{\xi}{1-\xi^2}\widetilde{\mathcal{E}}_T \right] ,
\end{eqnarray}
\end{subequations}
The $f_6$ amplitude is 0 in this model, since the corresponding $g_6$ is zero, as seen in Eqn.~\ref{g_functions}.
With these amplitudes, all of the observables of Section IIA will contain bilinears in the MPFFs. The asymmetries will involve the interference between real and imaginary parts of the bilinear products.
%
%
  
\subsection{Model for Transverse GPDs}
\label{sec-3-model}
We performed calculations using a 
model for the chiral-odd GPDs derived 
from the parametrization of Refs.\cite{AHLT1,AHLT2} (AHLT).
The parameterization's form for the unpolarized GPD $H$ is

\[ H(X,\zeta,t) = G(X,\zeta,t) R(X,\zeta,t), \]

\noindent
where $R(X,\zeta,t)$ is a Regge motivated term that 
describes the low $X$ and $t$ behaviors, while the contribution of 
$G(X,\zeta,t)$, obtained using a spectator model, is centered at intermediate/large values 
of $X$:
\begin{equation}
\label{diq_zeta}
G(X,\zeta,t) = 
{\cal N} \frac{X}{1-X} \int d^2{\bf k}_\perp \frac{\phi(k^2,\lambda)}{D(X,\zeta,{\bf k}_\perp)}
\frac{\phi({k^{\prime \, 2},\lambda)}}{D(X,\zeta,{ \bf k}_\perp^\prime)}.   
\end{equation}
Here $k$ and $k^\prime$ are the initial and final quark momenta respectively (Fig.\ref{fig1}), 
$D(X,{\bf k}_\perp) \equiv  k^2 - m^2$, 
$D(X-\zeta/(1-\zeta),{\bf k}_\perp^\prime) \equiv  k^{\prime \, 2} - m^2$,%
${\bf k}_\perp^\prime = {\bf k}_\perp - (1-X)/(1-\zeta)\Delta_\perp$, 
$m$ being the struck quark mass, 
$\Delta=P-P^\prime$ being the four-momentum transfer, and: 
\begin{eqnarray}
k^2 & = & X M^2 - \frac{X}{1-X} M_X^{2}  - \frac{{\bf k}_\perp^2}{1-X} \\
k^{\prime \, 2} & = & \frac{X-\zeta}{1-\zeta} M^2 - 
\frac{X-\zeta}{1-X} M_X^{2} - \left({\bf k}_\perp - \frac{1-X}{1-\zeta} \Delta_\perp \right)^2 \frac{1-\zeta}{1-X},
\end{eqnarray}
with $M$, the proton mass, and $M_X$ the (flavor-dependent) diquark mass (we suppress the flavor
indices for simplicity), $\phi(k^2,\lambda)$ defines the  vertex functions in both the
scalar and axial-vector cases \cite{AHLT1}.  
The normalization factor includes the nucleon-quark-diquark coupling, and it
is set to ${\cal N} = 1$ GeV$^6$. $G(X,\zeta,t)$ reduces to 
the form given in Ref.\cite{AHLT1} in the $\zeta \rightarrow 0$ case.
Similar equations were obtained for the spin flip GPD $E$.

Similarly to  \cite{VandH,DieKro} the 
$\zeta=0$ behavior is constrained by enforcing both the forward limit:
\begin{equation}
H^q(X,0,0)  =  q_{val}(X), 
\label{qpdf}
\end{equation}
where $q(X)$ is the valence quarks distribution, and  
the following relations:
\begin{subequations}
\begin{eqnarray}
\int_0^1 dX H^q(X,\zeta,t) & = & F_1^q(t) 
\\ 
\int_0^1 dX E^q(X,\zeta,t) & = & F_2^q(t),  
\end{eqnarray}
\label{FF}
\end{subequations}
\noindent
which defines the connection with the quark's contribution to the Dirac 
and Pauli form factors.
The proton and neutron form factors are obtained as:
\begin{subequations}
\begin{eqnarray}
F_{1(2)}^p(t) & = & \frac{2}{3}F_{1(2)}^u(t) - \frac{1}{3}F_{1(2)}^d(t) + \frac{1}{3}F_{1(2)}^s(t) \\ 
F_{1(2)}^n(t) & = & -\frac{1}{3}F_{1(2)}^u(t)+  \frac{2}{3}F_{1(2)}^d(t) + \frac{1}{3}F_{1(2)}^s(t),
\end{eqnarray}
\label{FF2}
\end{subequations}
Notice that differently from  \cite{VandH,DieKro} the AHLT parametrization does not make use of a
``profile function'' for the parton distributions, defined as: 

\[ H(X,0,t) = q(X) \exp-[t f(X)],  \]
   
\noindent 
whereas in our case the forward limit, $H(X,0,0) \equiv q(X)$, 
is enforced non trivially.
In other words, with the effort of simultaneously having to provide 
a new parametrization of the PDFs at low initial scale, we gain 
both the flexibility and insight that are necessary to model the behavior 
at $\zeta, \, t \neq 0$. 
  
The $\zeta$-dependent constraints are given by the higher moments of GPDs. 
The $n=1,2,3$ moments of the NS combinations: $H^{u-d} = H^u-H^d$, and $E^{u-d} = E^u-E^d$ 
are available
from lattice QCD \cite{zan_talk,haeg}, $n=1$ corresponding to the nucleon 
form factors. In a recent analysis a parametrization was devised that takes into 
account all of the above constraints. The parametrization gives an excellent description 
of recent Jefferson Lab data in the valence region, namely at $\zeta=0.36$.  


The connection to the chiral-odd GPDs is carried out by considering the 
following ansatz, similarly to what was adopted in Ref.\cite{Anselmino} for the 
transversity distribution, $h_1(X) \equiv H_T(X,0,0)$
\begin{eqnarray}
H_T^q(X,\zeta,t) & = & \delta q H^{q, val}(X,\zeta,t) \\
\overline{E}_T^q & \equiv & 2 \widetilde{H}_T + E_T = \kappa_T^q H_T(X,\zeta,t)      
\end{eqnarray}
where $\delta q$ is the tensor charge, and $\kappa_T^q$ is the transverse anomalous
moment  introduced, and connected to the transverse component of the total angular momentum
in \cite{Bur2,Bur3}. 

\section{$Q^2$ dependence} 
\label{sec-5}
The $Q^2$ dependence for $\pi^o$  electroproduction off a proton target, 
according to the factorization hypothesis, resides in the hard subprocess  $\gamma^* q \rightarrow \pi^o q^\prime$, both in kinematical factors 
(see Eqs.(\ref{g_functions})), and in the description of the pion vertex.  In determining the latter,
particular care needs to be taken of the chiral odd nature  of the reaction outlined in Section \ref{subsec-2.2}.
This requires the pion wave function 
to be proportional to $\gamma_5$. In this way one obtains a chiral odd structure for the hard scattering amplitude as follows
\begin{eqnarray}
\label{sub_cs}
\gamma_5 (\!\not{k} + \!\not{q}) \gamma^\mu & = & (k_\nu+q_\nu)\frac{\gamma_5}{2} \left( \ \left[ \gamma^\nu,\gamma^\mu \right] +
  \left\{ \gamma^\nu,\gamma^\mu \right\} \right) = 
   (k_\nu+q_\nu)\gamma_5 \left( i \sigma^{\mu\nu} + g^{\mu\nu} \right)  \nonumber \\
 & \propto  &  i  \gamma_5 \sigma^{\mu\nu}
\end{eqnarray}
 $\pi^o$ production therefore singles out the proton's chiral odd structure.
 
 An important point made here  is that  the Lorentz structure of the process has to be taken into account
 in addition to  the structure of the pion vertex.
 By considering both parity and C-parity conservation, 
 and by making use of duality  one can view the $\gamma^* q \rightarrow \pi^o q^\prime$ reaction as a transition between 
 a vector particle  ($\gamma^*$),  and either a vector or an axial vector particle (the two quark legs in Fig.\ref{fig1}a), with the 
 emission of a $\pi^o$.  
 
 Notice  that had one used a $\gamma^\mu \gamma_5$ term at the pion vertex, 
 based on the observation that  this is the only collinear, leading twist  
 contribution dictated by  the Operator Product Expansion (OPE), one would have 
 obtained $J^{PC}=1^{++}$, t-channel quantum numbers, 
 and a clear violation of $C$-parity
 in the  $\gamma^* q \rightarrow \pi^o q^\prime$ reaction would ensue. 
 Our procedure is to model the pion wave function in this process 
 after imposing $C$-parity conservation.
 This poses the problem of going beyond the
 collinear OPE-motivated description, an issue discussed also in \cite{JP}.
 
Here we propose a new model, using crossing symmetry and duality, in which we replace the calculation of the hard subprocess 
amplitudes (given in Eqs.(\ref{g_functions})),  that exhibit the structure in Eq.(\ref{sub_cs}),
with the $\gamma^*$-axial/vector meson-$\pi^0 \approx  \gamma^* (q \bar{q})  \rightarrow \pi^0$ vertex. 
This allows us to introduce OAM in our model.
%
%
A similar structure can be considered both for the GPD and Regge based descriptions.
For the latter, at non-zero $Q^2$ the kinematics shifts, so that $s$ depends on $Q^2$ 
and $x_{Bj}$, the preferred variables for the exclusive process. This kinematic generalization alters the Regge 
behavior vs. $Q^2$ and $t$ from the real photon limit. 
But furthermore there is a very strong suppression of the amplitudes and 
cross sections for increasing photon virtuality. 
This is indicated by data from DESY \cite{Hermes_rho,Hera_rho} and by 
theoretical expectations~\cite{BrodskyLepage} that for large $Q^2$ the amplitudes approach dimensional 
counting 
requirements from QCD, which predict $1/Q^4$. 
However, the transition from low to high $Q^2$ is subject to interpretation. Furthermore the data of 
interest for $\pi^0$ production at Jlab and Compass kinematics are at relatively low values of $Q^2$. 

%
 
Details of our model will be given in a forthcoming paper~\cite{GolLiu}.
Here we notice that the quantum numbers of the desired transition form factors 
can be identified with the ones for the following $J=1$ mesons (Fig.\ref{fig2}), 

\[ \gamma^* \rho (\omega) \pi^o \] 

\[ \gamma^* b_1 (h_1) \pi^o, \]   
or with  isovector (isoscalar) vector and axial vector exchanges, respectively.  

%
The quark contents of both vertices are:
\footnote{In principle a strange quark component appears in $h_1$. This 
can be, however, disregarded due to the small contribution to the nucleon strange
structure function.} 
\[
\rho \;  (b_1)  \rightarrow u \bar{u} - d \bar{d} \]
\[
\omega \;  (h_1)  \rightarrow  u \bar{u} + d \bar{d} \]
In the transition between the vector mesons $J^{PC} = 1^{--}$, and the $\pi^o$, $J^{PC} = 0^{-+}$,
the quark-antiquark pair carries an Orbital Angular Momentum (OAM) of $L=0$, both in the initial and final state.   
The transition between the axial-vector mesons $J^{PC} = 1^{+-}$, and the $\pi^o$, 
is instead characterized by a change of OAM ($L=1 \rightarrow L=0$).

Both the vector (V) and axial-vector (A) vertices have the following Lorentz structure: 
\begin{equation}
\Gamma_\mu=-ie^2F_{\pi\gamma}(Q^2) K_\mu,
\end{equation}
where the covariant kinematic factor for the vector case is
\begin{equation}
K_\mu=\epsilon_{\mu\nu\rho\sigma}p^\nu q^\rho \epsilon^\sigma(q^\prime),
\label{Kmu}
\end{equation} 
and the index $\mu$ refers to the virtual photon ($q^2=-Q^2$), while the $\epsilon^\sigma$ refers to the real photon ($q^{\prime 2}=0$) or the vector meson. 
For an axial vector the general form has two form factors,
\begin{equation}
\epsilon_\mu(q)K^\mu=-ie^2 [F_{\pi A}^{(1)}(Q^2) \epsilon(q)\cdot\epsilon^\prime(q^\prime)+F_{\pi A}^{(2)}(Q^2) 
(\epsilon^{\prime\mu}(q^\prime) \epsilon^\nu(q) + \epsilon^{\prime\nu}(q^\prime) \epsilon^\mu(q)  + (\epsilon^{\prime}(q^\prime) \cdot \epsilon(q)) g^{\mu\nu})q_\mu q^\prime_\nu].
\label{bvertex}
\end{equation}
So, with no assumptions about the form factors, the longitudinal photon going to a transverse axial vector meson 
dominates the first, S-wave part (with a factor of $q^{\prime \perp}\nu$ in the Lab frame). For the second part 
the transverse to transverse transition carries a similar factor. Other transitions are suppressed. For the vector 
exchange we can see that the transverse to transverse dominates (with a factor of $\nu p_{long}$). The inclusion of 
the form factors into this mix leads to more complicated conclusions. 

The structures given above are equivalent to the ones obtained in Eqs.(\ref{g_functions}), where
the coupling  $g_\pi$ is now replaced by combinations of vector and axial vector form factors.
While this will be explicitly shown in  \cite{GolLiu}, 
here we state the essential result for the parameterization of the $Q^2$ dependence that axial vector exchange dominates the 
longitudinal photon amplitude. For the transverse photon both vector and axial vector will contribute.
This is valid in a partonic picture as well as in the Regge approach since it is 
based only on the $J^{PC}$ quantum numbers for the different processes.   
We summarize our formulation of the $Q^2$ dependence in Table I. 
\begin{table}[h]
\protect\label{table-1}
\center
\begin{tabular}{|c|c|c|c|}
\hline
\hline
 Photon polarization  & t-channel parity & t-channel C-parity & t-channel polarization     \\ 
\hline
\hline
 L &  A  & -1 & L  \\ \hline
 L &  V  & -1 & Not allowed  \\ \hline
 T & A   & -1 & L, T  \\ \hline
 T & V   & -1 & T   \\ \hline
\hline
\end{tabular}
%
\caption{Dominating transitions for the different t-channel exchanges in the reaction 
$\gamma^* p \rightarrow \pi^o p$.}
\end{table}
From the Table I
 it is clear that the following contributions from the $t$-channel spin/parity components 
will go into the helicity amplitudes $f_i$ ($i=1,6$) of interest,

\[ f_1 = f_4 \propto  V, \; f_2 \propto  A+V, \; f_3 \propto  A-V , \; f_5 \propto  A.  \] 

The transition form factors  can be expressed in a PQCD model with transverse configuration space variables as \cite{StermanLi,KroJak} 
\begin{eqnarray}
F_{\gamma^* V \pi^o} & = & \int dx_1 dy_1 \int d^2{\bf b}  
\, \psi_{V}(y_1,b) \, 
\mathcal{C} K_0(\sqrt{x_1(1-x_1)Q^2}b)  \, \psi_{\pi^o}(x_1,b) exp(-S)   
\end{eqnarray}
%
where $x_1(y_1)$ is the longitudinal momentum fraction carried by the quark, ${\bf b}$ is the Fourier transform of the transverse intrinsic momentum, ${\bf k_T}$,  $\psi_{V}(y_1,b)$ and $\psi_{\pi^o}(x_1,b)$ are the vector meson and pion wave functions in configuration space, respectively, $\mathcal{C} K_0$ is the Fourier transform of the hard scattering amplitude, where  $K_0$ is the modified Bessel function of order zero, $\mathcal{C} = 8 \alpha_S(x_1 y_1Q^2) C_F$, and 
$exp(-S)$ is the Sudakov exponential. 
It is important to observe that it is sufficient for our purpose to use the leading twist pion wave function since the power suppression 
due to the $\gamma^5$ coupling is already accounted for in $g_2$ and $g_5$ (Eqs.(\ref{g_functions})).   
Furthermore, in the vector case the OAM is the same in the initial and final states ($L=0$) whereas 
for an axial-vector meson in the initial state, the OAM changes from $L=1$ to $L=0$,  leading to
\begin{eqnarray}
F_{\gamma^* A \pi^o} & = & \int dx_1 dy_1 \int d^2{\bf b}  
\, \psi_{A}^{(1)}(y_1,b) \, 
\mathcal{C} K_o(\sqrt{x_1(1-x_1)Q^2}b)  \, \psi_{\pi^o}(x_1,b) exp(-S) 
\end{eqnarray}
where now
\begin{equation}
\psi_{A}^{(1)}(y_1,b) = \int d^2 k_T J_1(y_1 b) \psi(y_1,k_T),  
\end{equation}
a higher order Bessel function appears as a consequence of having $L=1$ in the initial state 
\cite{RCARnold}. 

In impact parameter space this yields 
configurations of larger radius. 
In terms of meson distribution amplitudes this is described by functions of higher twist originating
from the ``bad'' components of the quark spinors \cite{BalBra}.   
We evaluated the form factors by using the asymptotic twist-two, $\phi_L(T) = 6x(1-x)$, 
and twist-three, $g_T = (3/4)[1+(2x-1)]^2$, amplitudes, defined in Ref.~\cite{BalBraTan}, 
corresponding to the same isospin but different spin configurations for the two mesons. 

 

We conclude by noting that while our approach might shed some light on the presence of large
transverse polarization components in a number of recent exclusive measurements, 
we cannot straightforwardly apply it to
vector meson production, since this is dominated by t-channel exchanges other than the axial
and vector types governing $\pi^o$ production.   
In summary, we introduced a model for the $Q^2$ dependence for vector and axial vector exchanges. 
They differ because in the axial vector cases there is a change of one unit of OAM producing a 
suppression with respect to the vector.
More details and more comparisons with $\pi^o$ electroproduction data will be 
given in a forthcoming manuscript
\cite{GolLiu}.   

\section{Results}
\label{sec-6}
We now present our quantitative results for $\pi^o$ electroproduction cross sections and 
asymmetries both
in the kinematical regime of currently analyzed experimental data obtained 
at $\gamma^* p$ CM energy,  $4$ GeV$^2$ $ \lesssim W \lesssim$ $9$ GeV$^2$, 
and $Q^2$ in the multi-GeV region, and in a larger energy and momentum transfer regime. 
Approximate scaling was found to hold in the case of DVCS \cite{DVCSA,DVCSB}. 
We therefore expect our picture 
based on chiral odd GPDs to be valid in this regime.
A Regge type description can also be reliably applied in this region for $-t << s$.     
The interplay between the Regge and partonic descriptions 
is key to the physical interpretation of GPDs and TMDs, and it
constitutes the main motivation of our study.          
%
Measurements from Jefferson Lab on $\pi^0$ 
production \cite{HallA_proposal} show non-negligible, larger than theoretically surmised, 
contributions from transversely polarized photons. 
Important aspects of our approach that guided us towards an interpretation 
of $\pi^o$ electroproduction data are that: {\it i)} a multi-variable analysis needs to be performed that 
is sensitive to the values
of the tensor charges, $\delta u$ and $\delta d$, and of the transverse moments, $\kappa_T^u$ and 
$\kappa_T^d$; {\it ii)} we consider a different   
$Q^2$ dependence of natural and unnatural t-channel exchanges governing
both the Regge and GPD approaches. 
In what follows we provide a survey of the effects of the variations of the transversity parameters 
on different observables. 

\subsection{Cross Sections}
\begin{figure}
\includegraphics[width=7.5cm]{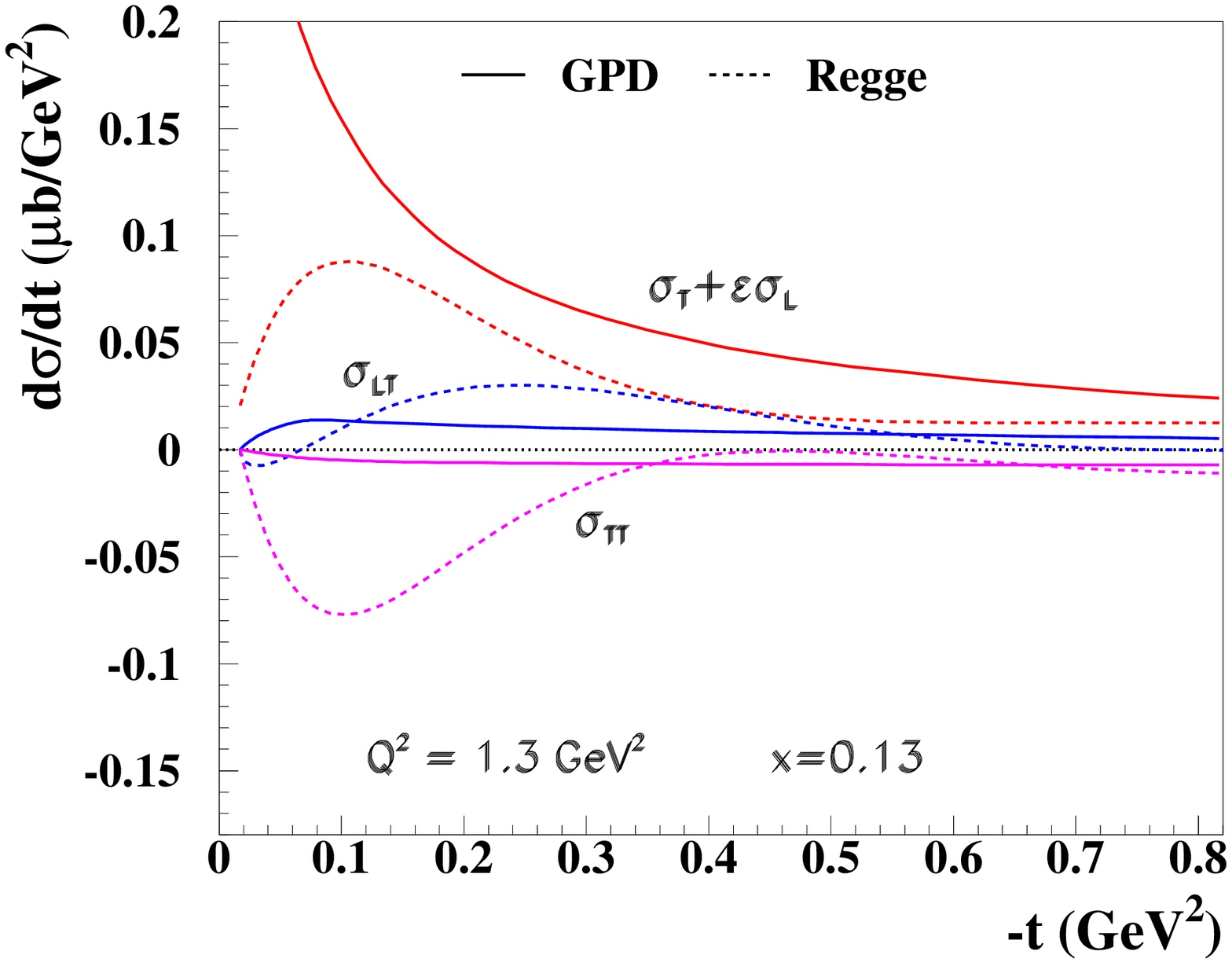}
\includegraphics[width=7.5cm]{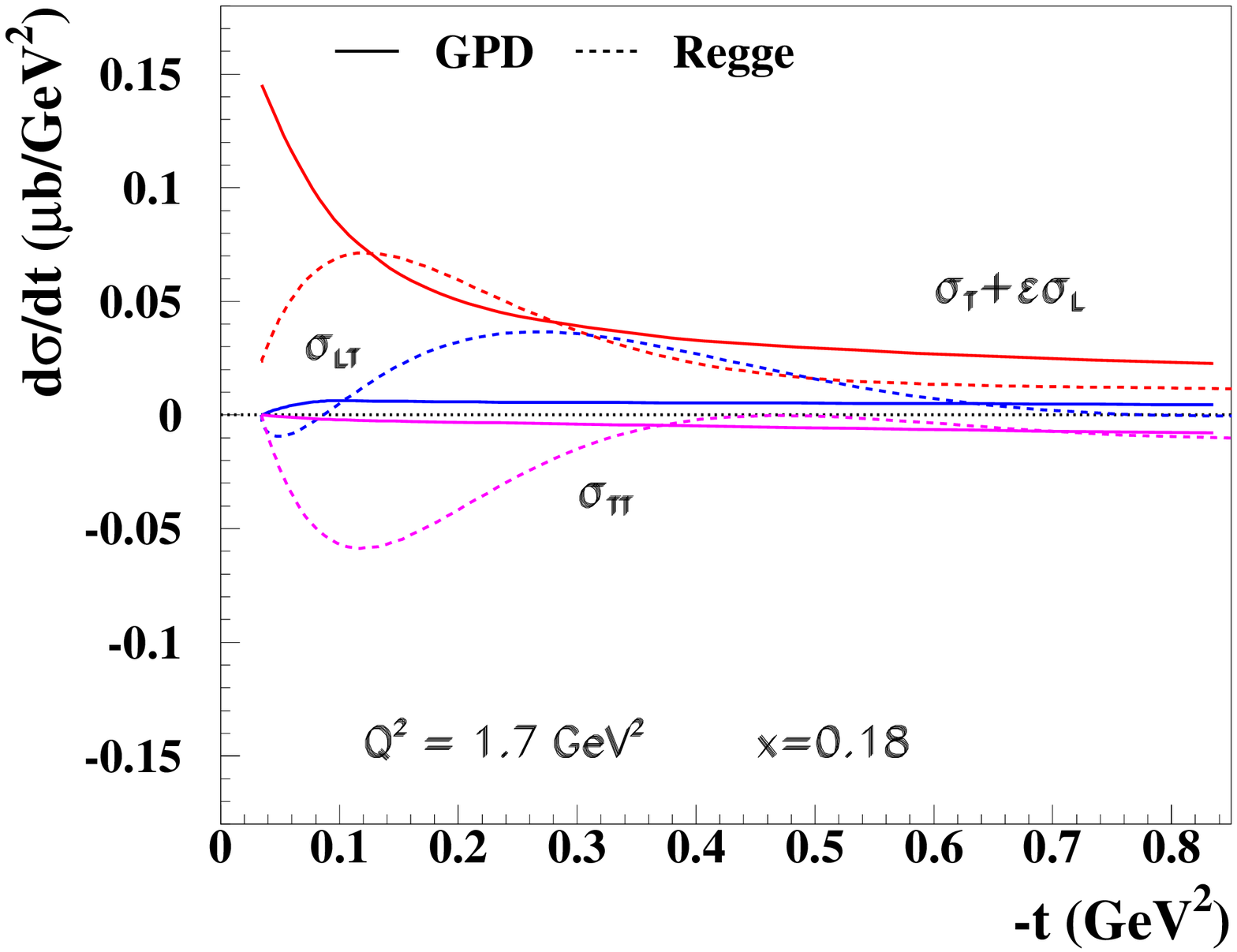}
\includegraphics[width=7.5cm]{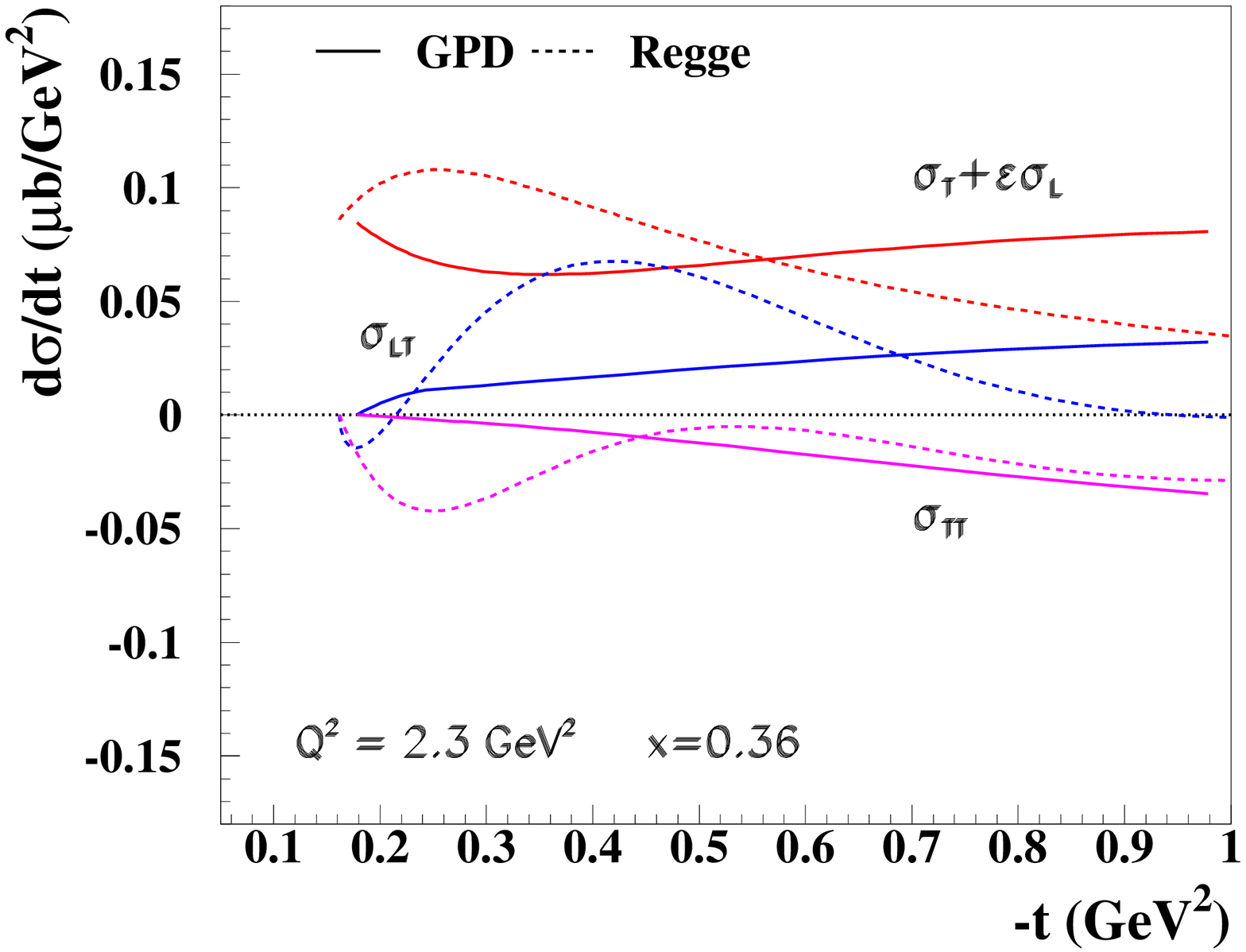}
\includegraphics[width=7.5cm]{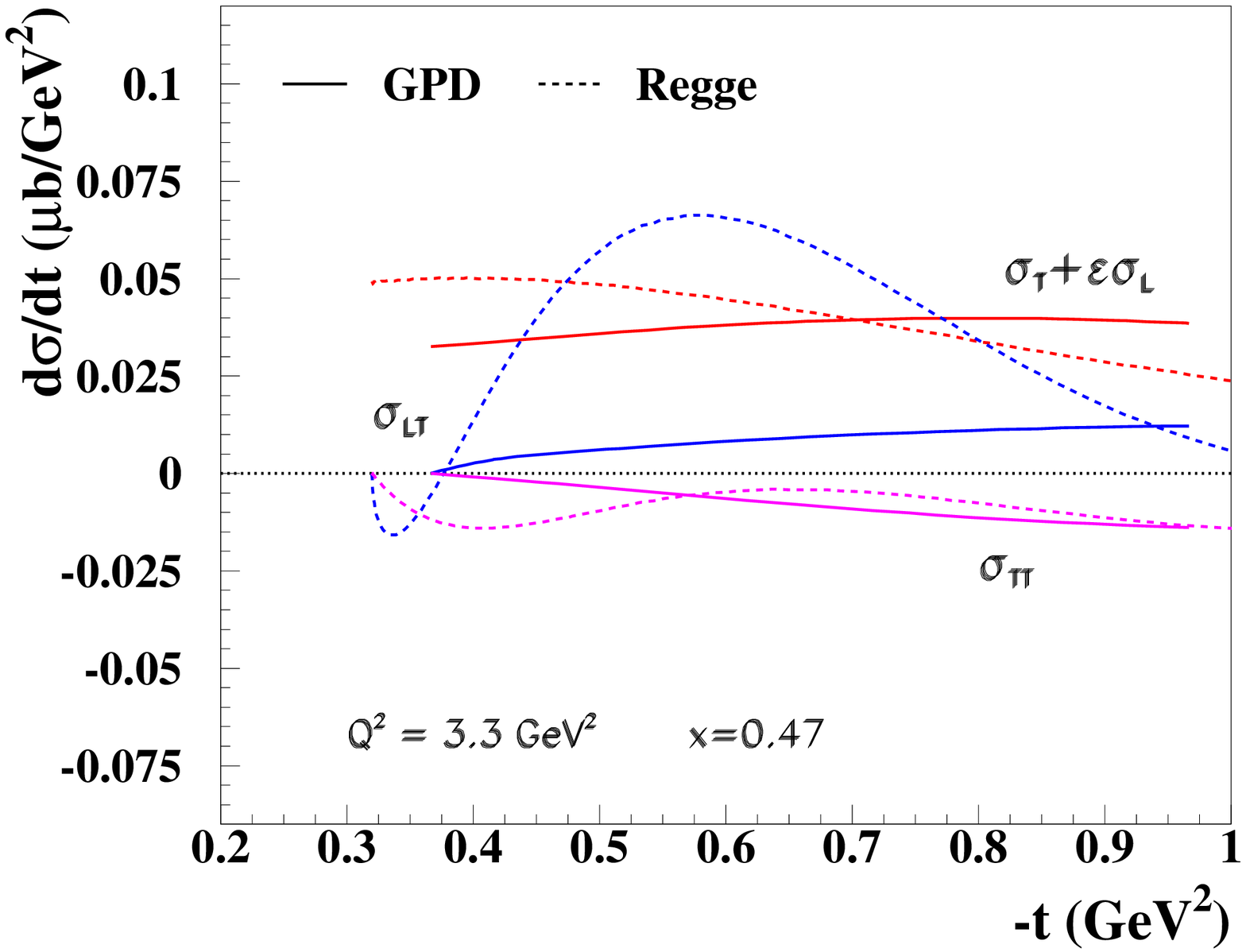}
\caption{(color online) Different contributions to the electroproduction 
cross section, Eq.(\ref{unp-xs}), plotted as a function of $-t$, 
in the Regge model (short dashes), and in the GPD model (full lines)
described in the text. Four different kinematical bins in $Q^2$ and $x_{Bj}$, in a range corresponding to recent
Jefferson Lab measurements are displayed. The parameters defining the tensor charges and the transverse anomalous
magnetic moments in the GPD model are respectively, $\delta u = 0.48$, $\delta d = -0.62$, {\it i.e.} consistent with the 
analysis of Ref.\cite{Anselmino}, and $\kappa_T^u=0.6$, $\kappa_T^d=0.3$.} 
\label{fig3} 
\end{figure}

\begin{figure}
\includegraphics[width=9.5cm]{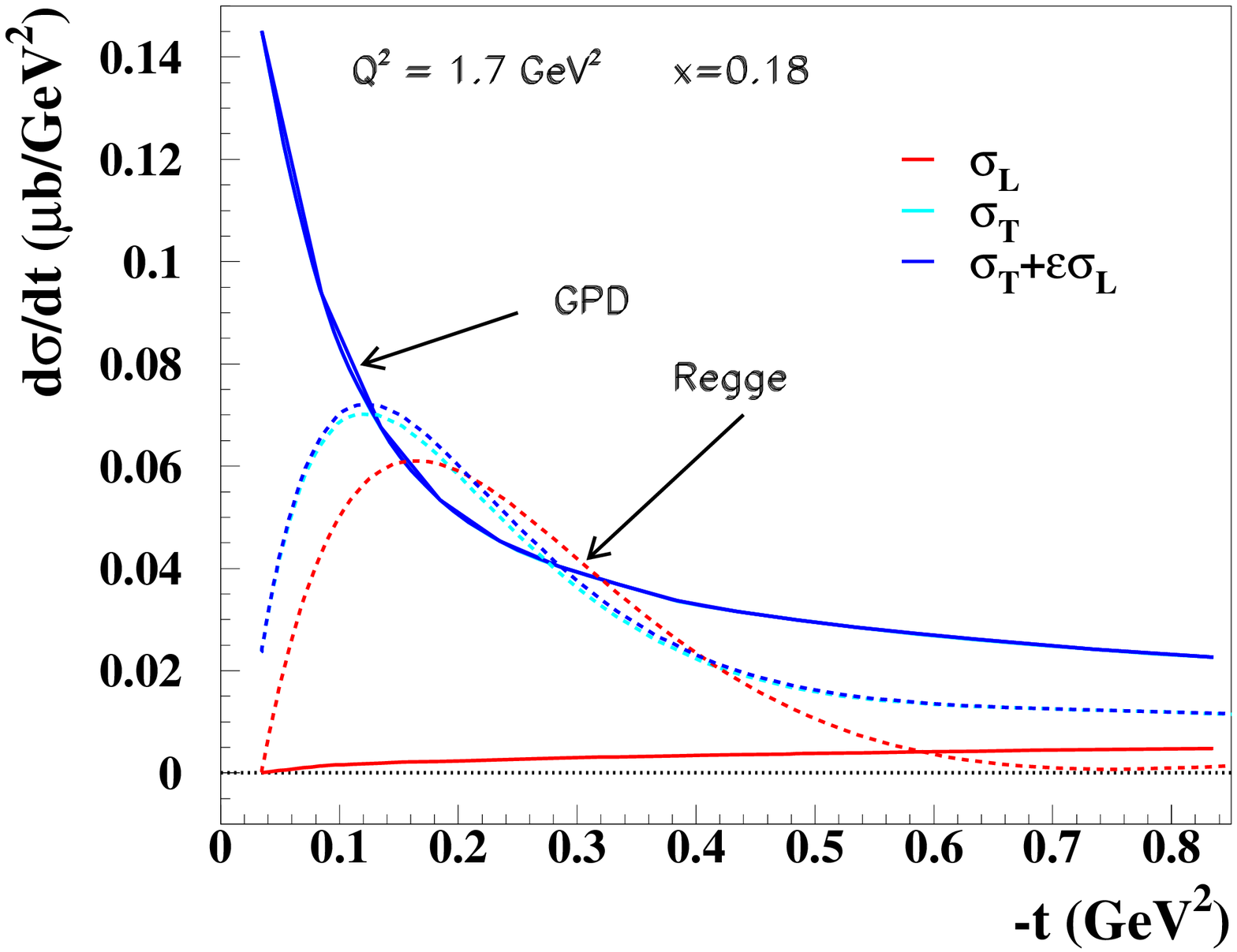} 

\includegraphics[width=9.5cm]{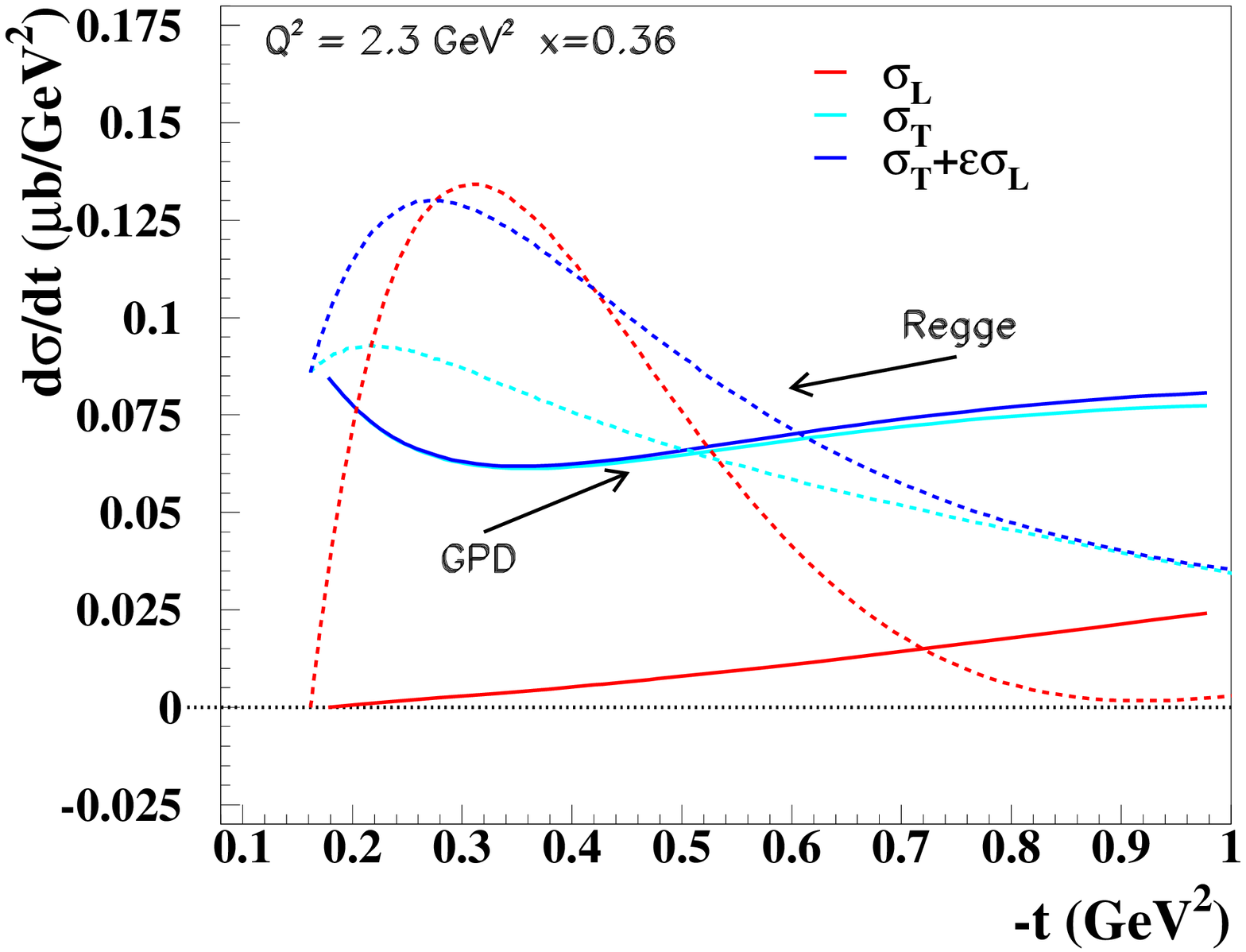}
\caption{(color online) Longitudinal, $\sigma_L \equiv d\sigma_L/dt$, Eq.(\ref{dsigL}), 
and transverse $\sigma_T \equiv d\sigma_T/dt$, Eq.(\ref{dsigT}) 
contributions presented along 
with their linear combination $\sigma_T+\epsilon_L \sigma_L$, in the  
electroproduction cross section, Eq.(\ref{unp-xs}), plotted as a function of $-t$. 
Both the Regge model (short dashes), and the GPD model (full lines)
are shown. Upper panel, $Q^2=1.7$ GeV$^2$, $x_{Bj}=0.17$; lower panel, $Q^2=2.3$ GeV$^2$, 
$x_{Bj}=0.36$. The parameters defining the tensor charges and the transverse anomalous
magnetic moments in the GPD model are respectively, $\delta u = 0.48$, $\delta d = -0.62$, 
and $\kappa_T^u=0.6$, $\kappa_T^d=0.3$.} 
\label{fig4} 
\end{figure}

\begin{figure}
\includegraphics[width=9.5cm]{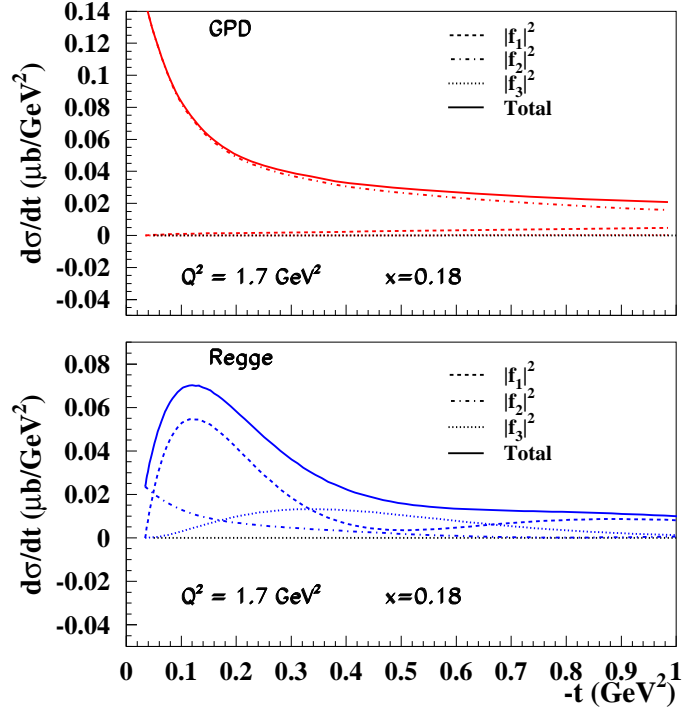}
\caption{(color online) Different helicity amplitudes contributions to the 
cross section, $\sigma = \sigma_T +\epsilon_L \sigma_L$, 
plotted vs. $-t$, at $Q^2=1.7$ GeV$^2$, $x_{Bj}=0.18$. Short dashes: $\mid f_1 \mid^2 \equiv \mid f_{1+,0+}\mid^2$;
dot-dashed line: $\mid f_2 \mid^2 \equiv \mid f_{1+,0-}\mid^2$; dotted line: 
$\mid f_3 \mid^2 \equiv \mid f_{1-,0+}\mid^2$  
(see Eq.(\ref{f1}) and following text). The full lines represent the total contributions. 
In the upper panel we show the GPD model, in the lower panel the
Regge model. Notice that the contribution of $f_3$ is very small in the GPD model at this kinematics.
The parameters defining the tensor charges and the transverse anomalous
magnetic moments in the GPD model are respectively, $\delta u = 0.48$, $\delta d = -0.62$, 
and $\kappa_T^u=0.6$, $\kappa_T^d=0.3$. An increase in $\kappa_T^{u,d}$ would produce larger
contributions of both $f_1$ and $f_3$ thus modifying the $t$-dependence of the cross sections.} 
\label{fig5} 
\end{figure}

In Figures \ref{fig3}, \ref{fig4} and \ref{fig5}, we show our predictions for the different contributions
to the $\gamma^* p \rightarrow \pi^o p$ cross section, obtained both in the Regge model, 
and in the GPD-based calculations. Both models predict similar trends in the measured 
experimental regime despite their seemingly different physical nature.
The different contributions to the cross section are in fact sensitive to the values 
of the tensor charge residing in the helicity 
amplitudes $f_2$ and $f_5$, as explained in the previous Sections, and defining their normalization
as $t \rightarrow t_{min}$. As for the 
helicity amplitudes $f_1$ and $f_4$, a connection can be established between GPDs and 
the (normalization of the) Boer-Mulders function through the concept of ``transverse spin 
anomalous magnetic moment'', $\kappa_T$. We expect a similar connection to 
be established for the Regge amplitudes as well. 
In the GPD model we used $\delta u =0.48$, $\delta d =-0.62$, namely the values of the 
tensor charges extracted from the global analysis of semi-inclusive data in Ref.\cite{Anselmino}, and 
the values $\kappa_T^u =0.6$, $\kappa_T^d=0.3$. The latter are smaller than currently available 
lattice Ref.\cite{haeg},  and model Ref.\cite{BofPas} calculations. It should be remarked that the $t$-dependence in the GPD model
follows closely what was found for the unpolarized case, {\it i.e.}  $\mathcal{H}$, in DVCS data in a similar
kinematical regime. The somewhat flatter $t$-dependence at large $x_{Bj}$ (lower panels in Fig.\ref{fig3}) 
is due to the interplay of 
the imaginary and real parts of the helicity amplitudes. In our approach, in fact, $\mathcal{H}_T$ 
has a similar trend to $\mathcal{H}$ in the unpolarized case. This is in turn determined by a parametrization constrained by the DVCS data. 
Both its real and imaginary parts 
are therefore decreasing with $-t$, but the real part being negative produces a less steep dependence 
of the cross sections with $-t$.     

\subsection{Asymmetries}
More marked differences between the two approaches appear 
both in the transverse target spin asymmetry, $A_{UT}$ and in the Beam Spin Asymmetry (BSA), 
proportional to $d \sigma_{LT^\prime}/dt$. 
In the GPD model used in this paper the size of
$A_{UT}$ in fact depends almost solely on the value of the tensor charge, due to the almost exact 
cancellation of the $\Im m (f_1^* f_3)$ term in Eq.(\ref{AUT}). $A_{UT}$ is therefore approximately 
proportional to $\mathcal{H}_T$. 
Such a cancellation does not occur in the Regge model, as it can be clearly seen at larger values 
of $-t$. This is a manifestation of the natural parity exchanges which
become dominant at larger $-t$.  
However, because, of the proportionality of the helicity amplitudes of $f_1$ and $f_3$ to $\sqrt{t_0-t}$ 
and $t_0-t$, respectively, as $t$ approaches $t_{min}$, the amplitude 
$f_2$, measuring the tensor charge is the main contribution at low $t$. 
This is consistent with the approximations used 
in our GPD model, and our proposed extractions are indeed valid at $t < Q^2$ where the GPD-based description of the electroproduction cross sections applies.    

On the other hand, in $d \sigma_{LT^\prime}/dt$, a cancellation occurs of the $\Im m (f_5^* f_2)$ 
term, Eq.(\ref{dsigLTp}). Therefore $d \sigma_{LT^\prime}/dt$ is only the dependent 
on the GPD, $\bar{E}_2$, allowing one in principle to measure the sensitivity to $\kappa_T$.  
Again, in the Regge model the above cancellation is only partial because of a more complicated 
interplay between the natural and unnatural parity exchanges. However we observe a similar 
trend showing the suppression of the tensor charge dependent term.  
We are therefore able to single out the observables $A_{UT}$ and $d \sigma_{LT^\prime}/dt$ as 
probes of the tensor charge and of $\kappa_T$.

\begin{figure}
\includegraphics[width=9.5cm]{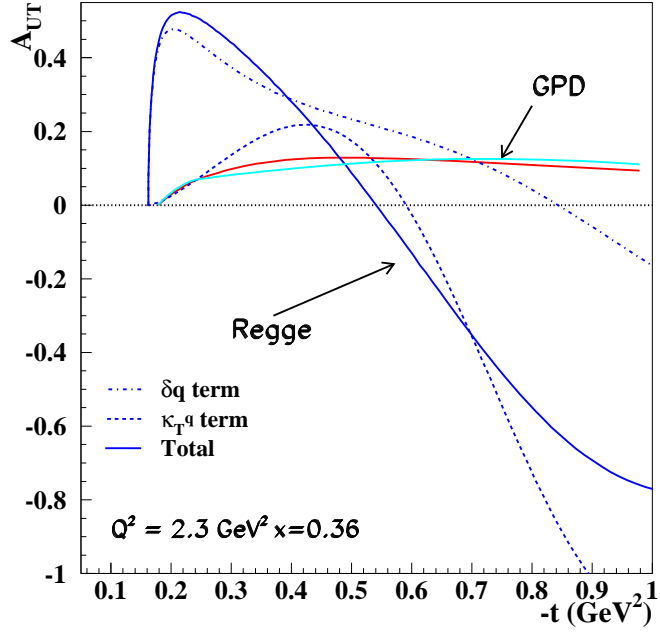}
\caption{(color online) The transverse spin asymmetry, $A_{UT}$, Eq.(\ref{AUT}) plotted vs. $-t$ in the
Regge model (short-dashes) and in the GPD model (full lines) 
$Q^2=2.3$ GeV$^2$, $x_{Bj}=0.36$. The GPD model 
is sensitive to the value of the u-quark and d-quark tensor charges taken here as $\delta u=0.48$, and
$\delta d = -0.62$, respectively. The rather small sensitivity to $\kappa_T^{u,d}$ is shown in the 
figure by plotting two different curves corresponding to:  $\kappa_T^{u}=0.6$, $\kappa_T^{2}=0.3$ and
$\kappa_T^{u}=3$, $\kappa_T^{2}=2$. For the Regge model we show separately 
the contributions of the combination
of helicities amplitudes including the tensor charge, namely $(f_4^*f_2)/\sigma_T$ (dot-dashes),  
and the ones sensitive to $\kappa_T^{u,d}$ only, $(f_1^*f_3)/\sigma_T$ (short-dashes).
The full line is the total contribution. One can see that the Regge model is dominated 
by the tensor charge contribution at low $-t$. In the GPD model the term $(f_1^*f_3)/\sigma_T$ in
Eq.(\ref{AUT}) cancels out exactly.} 
\label{fig6} 
\end{figure}

\begin{figure}
\includegraphics[width=9.5cm]{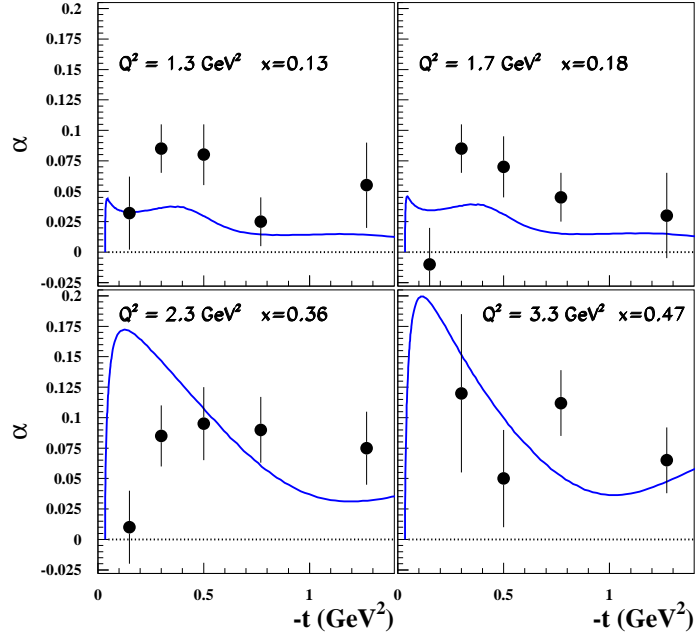}
\caption{(color online) Beam spin asymmetry parameter $\alpha$, Eq.(\ref{alpha}), plotted vs. $-t$ in the
Regge model at $Q^2=1.7$ GeV$^2$,  
$x_{Bj}=0.18$, $Q^2=2.3$ GeV$^2$, $x_{Bj}=0.36$, $Q^2=1.3$ GeV$^2$,  
$x_{Bj}=0.13$, and $Q^2=3.3$ GeV$^2$,  $x_{Bj}=0.47$. Experimental data from Ref.\cite{demasi}.} 
\label{fig7} 
\end{figure}

\begin{figure}
\includegraphics[width=9.5cm]{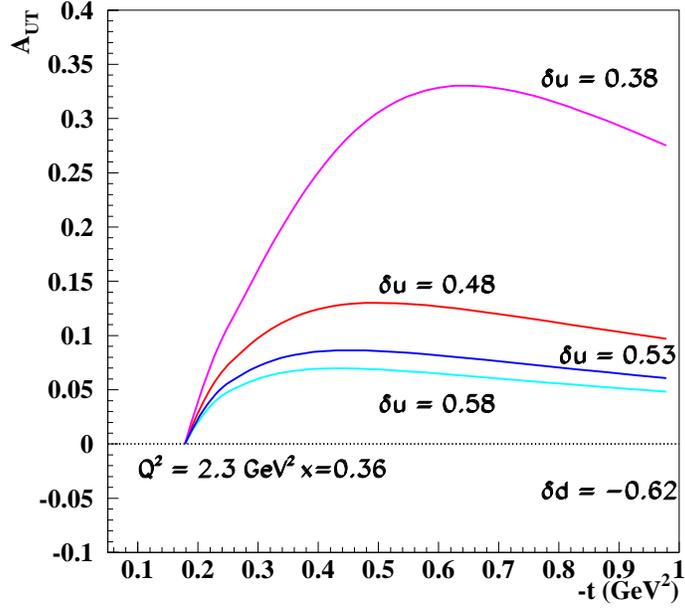}
\caption{(color online) Transverse spin asymmetry, $A_{UT}$, Eq.(\ref{AUT}), plotted vs. $-t$, at 
$Q^2=2.3$ GeV$^2$, $x_{Bj}=0.36$ for different
values of the $u$ quarks tensor charge, $\delta u$, used as a freely varying parameter in the GPD approach.
The $d$ quark component, $\delta d$ was taken as  $\delta d=-0.62$, {\it i.e.} equal to the central 
value extracted in the global fit of Ref.\cite{Anselmino}.} 
\label{fig8} 
\end{figure}

\begin{figure}
\includegraphics[width=9.5cm]{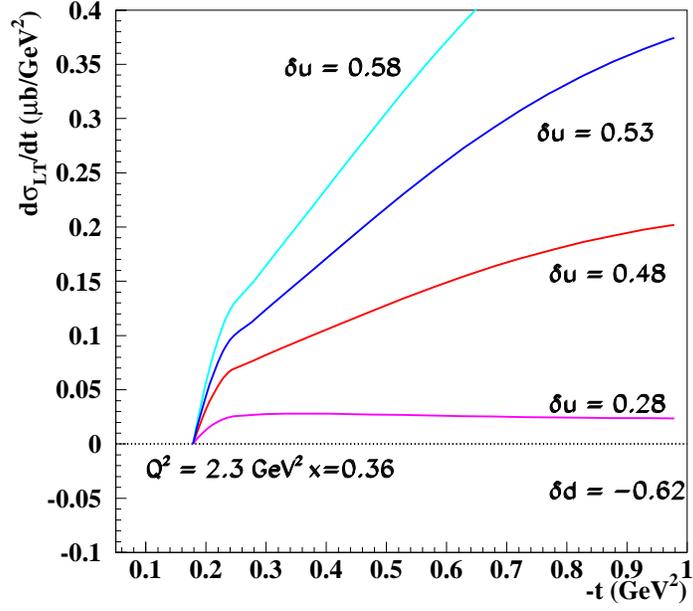}
\caption{(color online) Longitudinal/transverse interference term, $d \sigma_{LT}/dt$, Eq.(\ref{dsigLT}), 
plotted vs. $-t$ at $Q^2=2.3$ GeV$^2$, $x_{Bj}=0.36$, 
for different
values of the $u$ quarks tensor charge, $\delta u$, used as a freely varying parameter in the GPD approach.
The $d$ quark component, $\delta d$ was taken as  $\delta d=-0.62$, {\it i.e.} equal to the central 
value extracted in the global fit of Ref.\cite{Anselmino}.} 
\label{fig9} 
\end{figure}

Results for the asymmetries are shown in Figures 
\ref{fig6}, \ref{fig7},\ref{fig8},\ref{fig9}. 
$A_{UT}$, Eq.(\ref{AUT}), and $\alpha$, Eq.(\ref{alpha}),    
are given as a function of $t$ in Figs.\ref{fig6} and \ref{fig7}, respectively.
In Fig.\ref{fig6} we compare the Regge and GPD models. For the Regge we show separately 
the contributions of the combination
of helicity amplitudes including the tensor charge, and the ones sensitive to $\kappa_T$, 
together with the total contribution. More specifically we separate out the
terms proportional to $\Im m (f_1^* f_3)$ and $\Im m (f_1^* f_2)$, respectively. The latter, given
by the dot-dashed curve, clearly dominates the asymmetry at low $t$. The GPD model is instead governed
entirely by the tensor charge term.   
A similar picture is obtained for $\alpha$, shown in Fig.\ref{fig7}, by inverting the role of the
tensor charge and $\kappa_T^{u,d}$ terms. However, here the only surviving term in our GPD model
is $\Im m (f_5^* f_3)$, which is very small due to the smallness of the helicity amplitude $f_3$.
The Regge model gives larger contributions for $f_3$. These are 
shown at different recently measured kinematics. 
  
The sensitivity of $A_{UT}$ in the GPD model to the values of the u-quark and d-quark tensor charges, 
is shown in Fig.\ref{fig8}. The values in the figure 
were taken by varying up to $20 \%$ the values of the tensor charge extracted from the global 
analysis of Ref., {\it i.e.}  $\delta u=0.48$ and $\delta d = -0.62$, keeping the transverse
anomalous magnetic moment values, 
$\kappa_T^u = 0.6$ and $\kappa_T^d = 0.3$.
Fig.\ref{fig8} is one of the main results of this paper: it summarizes our 
proposed method for extracting the tensor charge from $\pi^o$ electroproduction experiments. 
A practical extraction of the tensor charge can be obtained  
by noticing that for the asymmetry, as well as for other quantities evaluated in this paper  
such as $d \sigma_{TT}/dt$, and $d \sigma_{LT}/dt$ 
the tensor charges for the different isospin components might be treated
as parameters related to the normalization of $H_T$ ($d \sigma_{LT}/dt$ is plotted in Fig.\ref{fig9}). 
Therefore
our model can be used to constrain the range of values allowed by the data.

\begin{figure}
 \includegraphics[width=9.5cm]{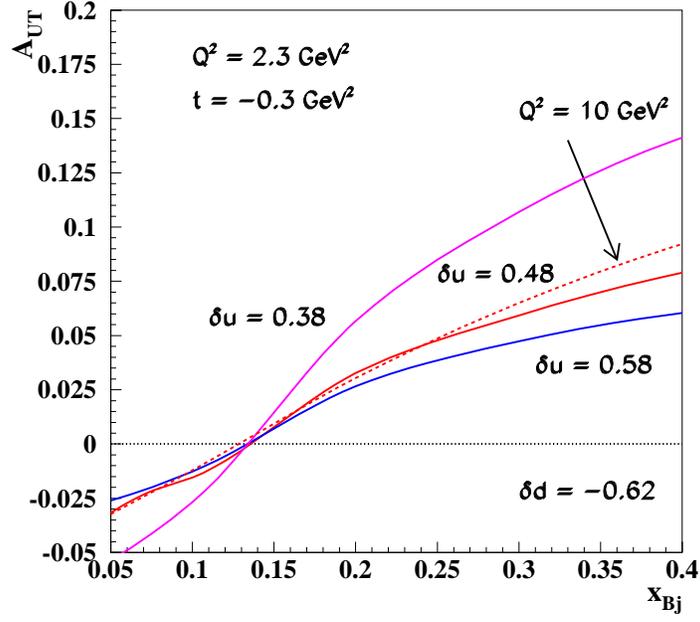}
\caption{color online) Transverse spin asymmetry, $A_{UT}$, Eq.(\ref{AUT}), plotted vs. $x_{Bj}$ at fixed $t=-0.3$ GeV$^2$, 
for different
values of the $u$ quarks tensor charge, $\delta u$ (notations as in Figures \ref{fig8} and \ref{fig9}).
$Q^2=2.3$ GeV$^2$ (full lines), 
Results at $Q^2=10$ GeV$^2$, for $\delta u =0.48$, $\delta d=-0.62$ are also shown 
(short dashed line).} 
\label{fig10} 
\end{figure}

\begin{figure}
\includegraphics[width=9.5cm]{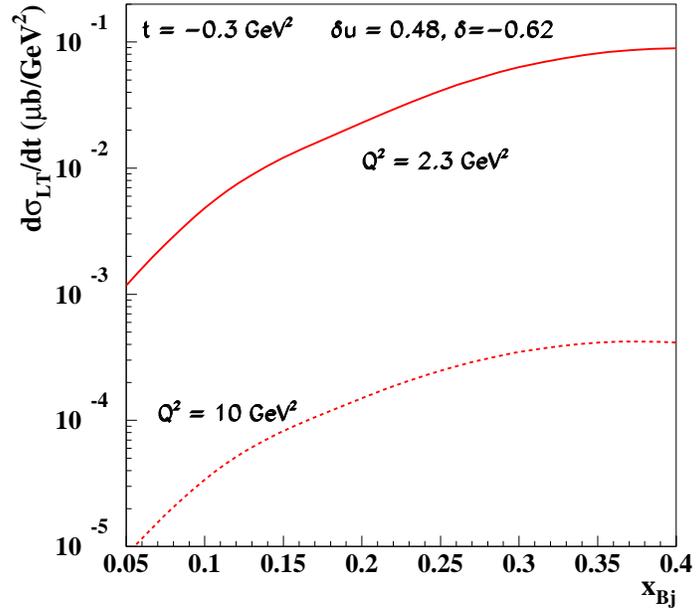}
\caption{(color online) $d \sigma_{LT}/dt$, plotted vs. $x_{Bj}$ at $t=0.3$ GeV$^2$ for two 
different values of $Q^2$: $Q^2=2.3$ GeV$^2$ (full line) and $Q^2=10$ GeV$^2$ (short dashed line).} 
\label{fig11} 
\end{figure}

In Fig.\ref{fig10} we show the sensitivity of $A_{UT}$ to the tensor charge values at fixed 
$t=-0.3$ GeV$^2$, and as a function of $x_{Bj}$. As in the previous figures we took 
$\kappa_T^u=0.6$ and $\kappa_T^d=0.3$. We performed calculations for a range of
values of $Q^2$. We find that the $Q^2$ dependence of $A_{UT}$ is rather small due to the
cancellations of the form factors in the ratio (Eq.(\ref{AUT})). On the contrary, as
can be seen from Fig.\ref{fig11} for {\it e.g.} $d\sigma_{LT}/dt$, 
the single contributions to the cross section expectedly display a steep $Q^2$ dependence.

\begin{figure}
\includegraphics[width=9.5cm]{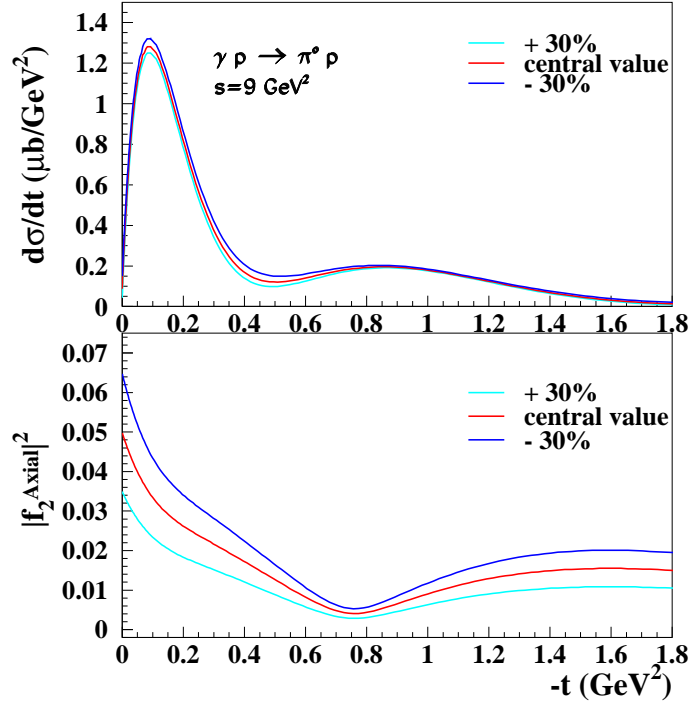}
\caption{(color online) Photoproduction cross section at $s=9$ GeV$^2$, calculated following
Ref.\cite{GolOwe} (upper panel).  The value of the tensor charge was extracted from the
Regge residue as described in the text (curve labeled as ``central''). Curves corresponding
to a $\pm 30 \%$ variation of the tensor charge are labeled correspondingly. In the lower panel we show
the axial vector contribution to the residue of the amplitude $f_2$. The tensor charge is extracted
from the value of the residue in $t \rightarrow 0$.}
\label{fig12} 
\end{figure}

Notice that the electroproduction data are essential in determining the tensor charge and other
transversity related quantities. This is illustrated in Fig.\ref{fig12} where we show
the photoproduction cross section calculated following the model in \cite{GolOwe}. 
The value of the tensor charge is extracted in this case from the
the Regge residue of the axial vector contribution to the helicity amplitude $f_2$ (see Section 
\ref{sec-3}) at $t \rightarrow 0$. This is plotted in the lower panel of Fig.\ref{fig12} where 
the central value
extracted by fitting the model parameters to the photoproduction data is shown along with
curves corresponding
to a $\pm 30 \%$ variation of the tensor charge (labeled correspondingly in the figure). 
From the figure it is clear that the photoproduction cross section is very little affected by variations 
in the values of the tensor charge, 
except very near forward.  

\subsection{$Q^2$ dependence}
\begin{figure}
\includegraphics[width=9.5cm]{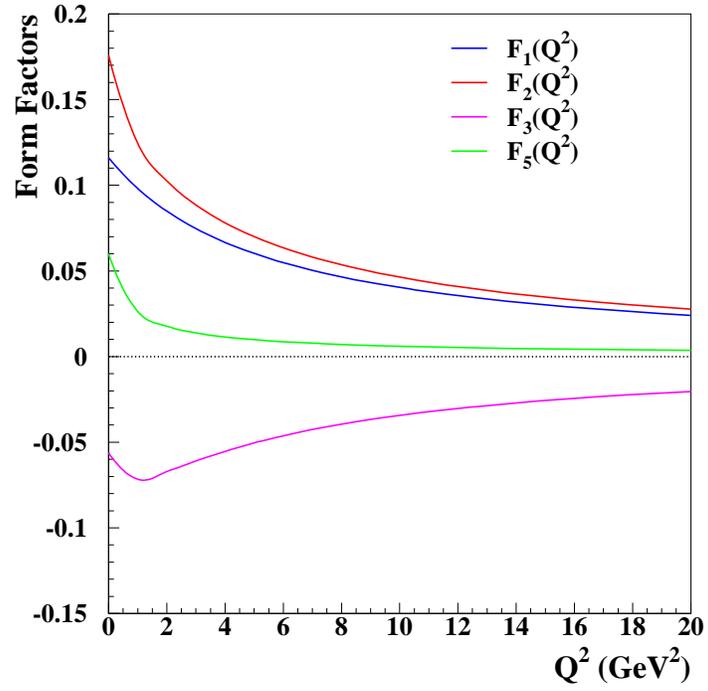}
\caption{(color online) Form factors describing the upper vertex in Fig.\ref{fig1}b,
plotted vs. $Q^2$ 
for the different helicity amplitudes, namely $F_1(Q^2) = F_V(Q^2)/2$, 
$F_2(Q^2) = (F_V(Q^2) + F_A(Q^2))/2$, $F_3(Q^2)= - (F_V(Q^2)-F_A(Q^2))/2$, and 
$F_5(Q^2)= F_A(Q^2))/2$, 
described in Section \ref{sec-5}. All form factors were calculated using the
approach described in Section \ref{sec-5}.} 
\label{fig13} 
\end{figure}

\begin{figure}
\includegraphics[width=9.5cm]{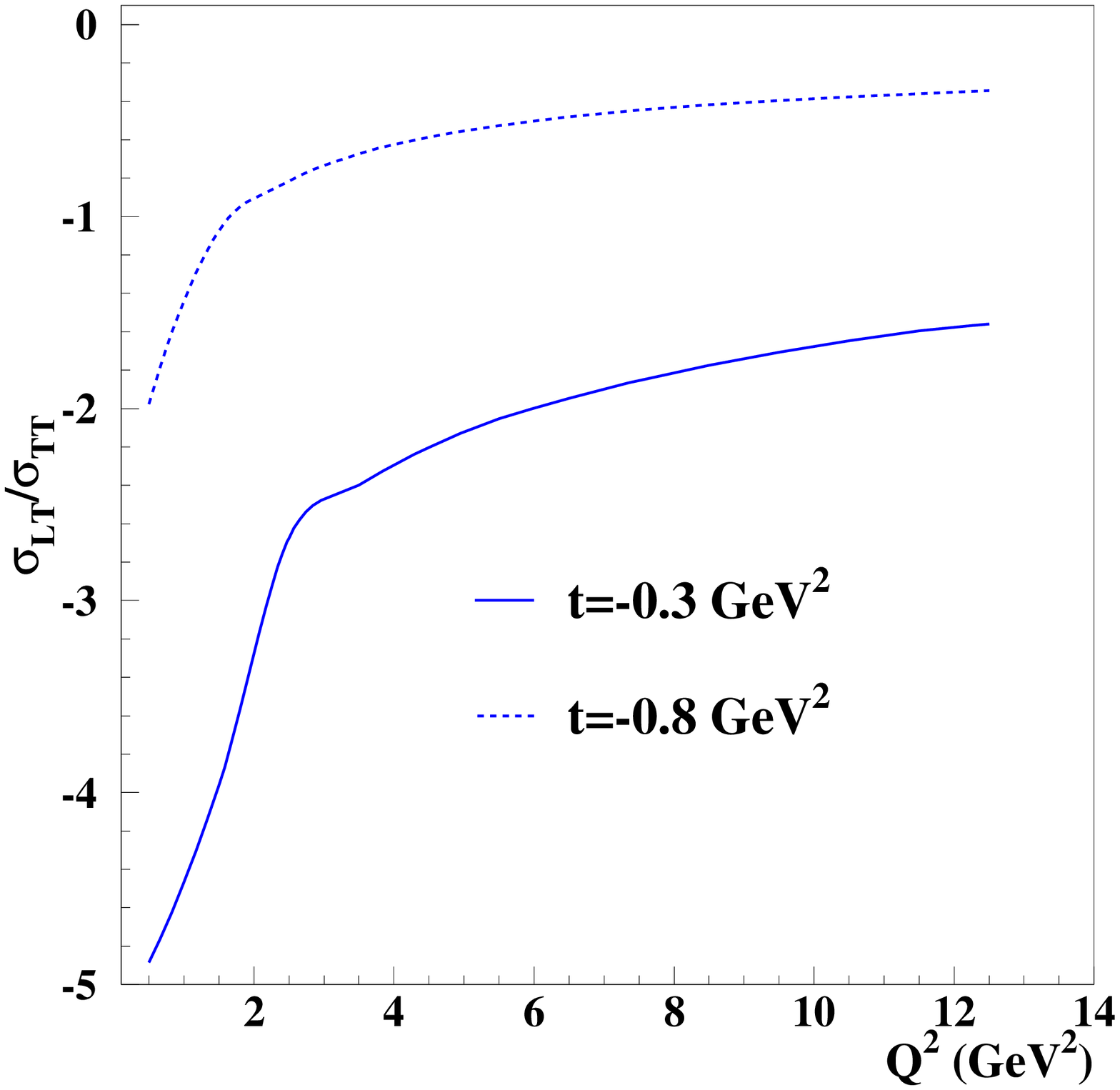}
\caption{Ratios $R= \sigma_{LT}/\sigma_{TT}$ plotted vs. $Q^2$, at $x-0.36$, 
for two different values of $t$, $t=0.3$ GeV$^2$, and $t=0.8$ GeV$^2$.} 
\label{fig14} 
\end{figure}

The $Q^2$ dependence of our model is illustrated in Figs. \ref{fig13} and \ref{fig14}. 
Fig.\ref{fig13} shows the different form factors describing the upper vertex in Fig.\ref{fig1}b,
for the different helicity amplitudes, namely $F_1(Q^2) = F_V(Q^2)/2$, 
$F_2(Q^2) = (F_V(Q^2) + F_A(Q^2))/2$, $F_3(Q^2)= - (F_V(Q^2)-F_A(Q^2))/2$, and $F_5(Q^2) = F_A(Q^2)/2$, where 
$F_V$ and $F_A$ are a short notation for the vector and axial transition form factors,$F_{\gamma^*V\pi}$, and
$F_{\gamma^*A\pi}$ introduced in Section \ref{sec-5}. All form factors were calculated using the
approach described in Section \ref{sec-5}. The axial form factor displays a steeper $Q^2$ dependence
at large $Q^2$, due to the difference in orbital angular momentum between the initial and final
hadronic states. In  Fig.~\ref{fig14} we show the impact of multiplying the different helicity amplitudes 
by different form factors on some of the observables which are governed by either longitudinal or 
transverse photon polarization. Results are shown at  $x_{Bj}=0.36$ for 
two different values of $t$, $t=-0.3$ GeV$^2$ and $t=-0.7$ GeV$^2$,
in the GPD model.
Despite the fact that the $t$ dependence plays an important role, as can be seen from Figs. \ref{fig3},\ref{fig4}, 
\ref{fig5}, we expect the longitudinal to transverse ratios,
$\sigma_L/\sigma_T \propto \left( F_A/(F_A+F_V) \right)^2$,
and $ \sigma_{LT}/\sigma_{TT} $, 
to have a less steep $Q^2$ dependence than the one based on simple PQCD predictions. 
The shape of the curves is a consequence of the difference in the $Q^2$ behavior for the axial and vector
form factors, whose ratio displays a $1/Q \log^a(Q^2/\Lambda)$ dependence. These however
enter the cross section in different linear combinations, and with different weights depending on the 
values of $t$, giving rise to the curves shown in the figure. It should be noticed that this is qualitatively
different from taking different monopole masses for axial vector and vector meson form factors, 
and assuming the same $Q^2$ behavior \cite{VdHLag}.

\section{Conclusions and Outlook}
\label{sec-7}
In conclusion, we presented a framework for analyzing $\pi^o$ exclusive electroproduction  
where, by observing that this reaction proceeds through 
C-parity odd and chiral odd combinations of t-channel exchange 
quantum numbers, it can be selected to obtain direct measurements of the meson production form factors
for the chiral odd generalized parton distributions. This is at variance with deeply virtual 
Compton scattering, and with both vector meson and charged $\pi$ electroproduction, 
where the axial charge corresponding to C-parity even exchanges can enter the amplitudes.
We then studied the different terms appearing in the 
cross section for scattering from an unpolarized proton, 
including the beam 
polarization asymmetry using the helicity amplitudes formalism. 

A Regge based description based on the ``weak cut'' approach was adopted where
the leading axial vectors exchanges, $b_1$ and $h_1$, determine 
the tensor charge and 
the transversity distribution, while the leading vector exchanges, $\rho$ and $\omega$
can be related to the transverse anomalous moment. 

The partonic description, singling out the chiral odd GPDs, was implemented to 
show the sensitivity of some of the observables, in particular the interference
terms in the unpolarized cross section to the values of the $u$ and $d$ quark 
tensor charges, as well as to the values of the $u$ and $d$ quark transverse anomalous 
moments.
Predictions were also given for the transverse target spin asymmetry, $A_{UT}$.   

The various correspondences between the Regge approach and the 
GPD models were highlighted. This aspect of the of analysis represents 
an avenue that we will continue to pursue in the near future.

Finally, we expect a variety of new flavor sensitive 
observables to be extracted from the data in the near 
future using both unpolarized data and asymmetries from transversely polarized
proton and deuteron data on $\pi^o$ and $\eta$ production at the higher $s$ 
values attainable at Jefferson Lab at 12 GeV. The extension of our analysis
to these types of reactions  promises to be a 
rich area for both theoretical and experimental exploration in the near future.

\acknowledgments
We thank Leonard Gamberg for participating in initial discussions. We are also
thankful to John Ralston, Paul Stoler, and Christian Weiss for useful comments. 
This work is supported by the U.S. Department
of Energy grants no. DE-FG02-01ER4120 (S.A. and S.L), and 
no. DE-FG02-92ER40702 (G.R.G.).

\appendix*
\section{Kinematical Transformations} 
\label{app-1}
We present the transformation from the pion's scattering angle in the laboratory frame
to the center-of-mass frame (CM):
\begin{equation}
d(\cos \theta^{LAB}_\pi ) = \frac{\gamma \left( 1 + \beta \cos \theta^{CM}_\pi \right)}%
{\left( \sin^2 \theta^{CM}_\pi + \gamma^2 \left(\cos \theta^{CM}_\pi +\beta \right)^2 \right)^{3/2} }
d (\cos \theta^{CM}), 
\end{equation}
with
\begin{subequations}
\begin{eqnarray}  
\gamma & = & \frac{\nu+M}{\sqrt{s}} \\
\beta & = & \frac{\sqrt{\nu^2 + Q^2}}{\nu+M},
\end{eqnarray}  
\end{subequations}
The expression in the kinematical invariant, $t$ is given by: 
\begin{equation}
d (\cos \theta^{CM}_\pi) = \left[(s+Q^2-M^2)(s-M^2)\right]^{-1/2} dt .
\end{equation}



\newpage

\begin{thebibliography}{99}
\bibitem{DMul1} D.~Muller, D.~Robaschik, B.~Geyer, F.~M.~Dittes and J.~Horejsi,
Fortsch.\ Phys.\  {\bf 42}, 101 (1994)

\bibitem{Ji1} X.~D.~Ji, Phys.\ Rev.\ D {\bf 55}, 7114 (1997)

\bibitem{Rad1} A.~V.~Radyushkin,
Phys.\ Rev.\ D {\bf 56}, 5524 (1997)

\bibitem{Bur} M.~Burkardt,
Int.\ J.\ Mod.\ Phys.\ A {\bf 18}, 173 (2003); {\it ibid}
Phys.\ Rev.\ D {\bf 62}, 071503 (2000)
[Erratum-ibid.\ D {\bf 66}, 119903 (2002)].

\bibitem{Die_rev} M.~Diehl,
Phys.\ Rept.\  {\bf 388}, 41 (2003).

\bibitem{BelRad} A.~V.~Belitsky and A.~V.~Radyushkin,
  Phys.\ Rept.\  {\bf 418}, 1 (2005)

\bibitem{BofPas} S.~Boffi and B.~Pasquini,
 arXiv:0711.2625 [hep-ph].
  
\bibitem{Diehl_01} M. Diehl, Eur. Phys. Jour. C {\bf 19}, 485 (2001).

\bibitem{JafJi}
R.~L.~Jaffe and X.~D.~Ji,
  Nucl.\ Phys.\  B {\bf 375}, 527 (1992).

\bibitem{Bur2} M. Burkardt, Phys. Rev. D {\bf 72}, 094020 (2005); {\it ibid}
Phys. Lett. {\bf B} 639 (2006) 462.

\bibitem{Bur3} M.~Burkardt and G.~Schnell,
  Phys.\ Rev.\  D {\bf 74}, 013002 (2006)
 
\bibitem{DieHag} M.~Diehl and Ph.~Hagler,
  Eur.\ Phys.\ J.\ C {\bf 44}, 87 (2005)
  [arXiv:hep-ph/0504175].

\bibitem{Metz} S. Meissner, A. Metz and K. Goeke, Phys. Rev. D {\bf 76}, 034002 (2007).  

 
\bibitem{BogMul} M.~Boglione and P.~J.~Mulders,
Phys.\ Rev.\  D {\bf 60}, 054007 (1999)

%
\bibitem{GolOwe}G.~R.~Goldstein and J.~F.~Owens,
  Phys.\ Rev.\  D {\bf 7} (1973) 865.

\bibitem{AHLT1} S.~Ahmad, H.~Honkanen, S.~Liuti and S.~K.~Taneja,
  Phys.\ Rev.\  D {\bf 75}, 094003 (2007). 

\bibitem{AHLT2} S.~Ahmad, H.~Honkanen, S.~Liuti and S.~K.~Taneja,
arXiv:0708.0268. 


\bibitem{GG1} L.~P.~Gamberg and G.~R.~Goldstein,
  Phys.\ Rev.\ Lett.\  {\bf 87}, 242001 (2001)
  
 
\bibitem{CFS} J.C. Collins, L. Frankfurt and M. Strikman, 
 Phys.\ Rev.\  D {\bf 56}, 2982 (1997)

\bibitem{BroFraGun}S.~J.~Brodsky, L.~Frankfurt, J.~F.~Gunion, A.~H.~Mueller and M.~Strikman,
 Phys.\ Rev.\  D {\bf 50}, 3134 (1994)
 
\bibitem{CF} J.~C.~Collins and A.~Freund,
  Phys.\ Rev.\  D {\bf 59}, 074009 (1999)

\bibitem{Mank} L.~Mankiewicz, G.~Piller and A.~Radyushkin,
  Eur.\ Phys.\ J.\  C {\bf 10}, 307 (1999)

\bibitem{AniTer} I.~V.~Anikin and O.~V.~Teryaev,
    Phys.\ Lett.\  B {\bf 554}, 51 (2003)
 
\bibitem{Pire} M. Diehl, T. Gousset and B. Pire, Phys. Rev. D {\bf 59}, 034023 (1999) 

\bibitem{Pire2} D.Y. Ivanov, {\it et al.}, Phys. Lett. B {\bf 550}, 65 (2002)
 
\bibitem{DieVin} M.~Diehl and A.~V.~Vinnikov,
  Phys.\ Lett.\  B {\bf 609}, 286 (2005)
  
\bibitem{HuaKro} H.~W.~Huang and P.~Kroll,
  Eur.\ Phys.\ J.\  C {\bf 17}, 423 (2000)
 
\bibitem{SzcLlaLon} A.~P.~Szczepaniak, J.~T.~Londergan and F.~J.~Llanes-Estrada,
   arXiv:0707.1239 [hep-ph].

\bibitem{Hermes_rho}  
 A.~Airapetian {\it et al.}  [HERMES Collaboration],
  Phys.\ Rev.\ Lett.\  {\bf 87}, 182001 (2001).

\bibitem{Hera_rho} S.~Chekanov {\it et al.}  [ZEUS Collaboration],
  PMC Phys.\  A {\bf 1}, 6 (2007);   V.~Y.~Alexakhin {\it et al.}  [COMPASS Collaboration],
 Eur.\ Phys.\ J.\  C {\bf 52}, 255 (2007); A.~Airapetian {\it et al.}  [HERMES Collaboration],
  Eur.\ Phys.\ J.\  C {\bf 17}, 389 (2000); 
S.~Aid {\it et al.}  [H1 Collaboration],
Nucl.\ Phys.\  B {\bf 468}, 3 (1996).

\bibitem{VGG} M.~Vanderhaeghen, P.~A.~M.~Guichon and M.~Guidal,
  Phys.\ Rev.\  D {\bf 60}, 094017 (1999)

\bibitem{BelJiYuan_F2} Belitsky, X. Ji, and F. Yuan, Phys. Rev. Lett. {\bf 91}, 092003 (2003). 

\bibitem{BroHil} S.~J.~Brodsky, J.~R.~Hiller, D.~S.~Hwang and V.~A.~Karmanov,
  Phys.\ Rev.\  D {\bf 69}, 076001 (2004).
  
\bibitem{JP} T. Gousset, B. Pire and J.P. Ralston, Phys. Rev. D {\bf 53}, 1202 (1996). 

\bibitem{kubarovsky} V.~Kubarovsky, P.~Stoler, I.~Bedlinsky and f.~t.~C.~Collaboration,
  arXiv:0802.1678 [hep-ex]. 

\bibitem{RasDon} A.~S.~Raskin and T.~W.~Donnelly,
  Annals Phys.\  {\bf 191}, 78 (1989)
  [Erratum-ibid.\  {\bf 197}, 202 (1990)].

\bibitem{GolMor} G.~R.~Goldstein and M.~J.~Moravcsik,
  Int.\ J.\ Mod.\ Phys.\  A {\bf 1}, 211 (1986).
  
\bibitem{GolKro} S.~V.~Goloskokov and P.~Kroll,
  Eur.\ Phys.\ J.\  C {\bf 50}, 829 (2007).
  
\bibitem{Martinetal} A.D. Martin, M.G. Ryskin and T.Teubner, 
  Phys.\ Rev.\  D {\bf 55}, 4329 (1997)

\bibitem{HoyLen} P.~Hoyer, J.~T.~Lenaghan, K.~Tuominen and C.~Vogt,
  Phys.\ Rev.\  D {\bf 70}, 014001 (2004)

\bibitem{Ivanov} I.~F.~Ginzburg and D.~Y.~Ivanov,
  Phys.\ Rev.\  D {\bf 54}, 5523 (1996)

\bibitem{RalSop} J.~P.~Ralston and D.~E.~Soper,
  Nucl.\ Phys.\  B {\bf 172}, 445 (1980).
 
\bibitem{VandH} M.~Guidal, M.~V.~Polyakov, A.~V.~Radyushkin and M.~Vanderhaeghen,
  Phys.\ Rev.\ D {\bf 72}, 054013 (2005).

\bibitem{DieKro} M.~Diehl, T.~Feldmann, R.~Jakob and P.~Kroll,
  Eur.\ Phys.\ J.\ C {\bf 39}, 1 (2005).

\bibitem{zan_talk} M.~Gockeler {\it et al.}  [QCDSF Collaboration],
  PoS {\bf LAT2007}, 147 (2007)
  [PoS {\bf LATTICE2007}, 147 (2006)]
  [arXiv:0710.2489 [hep-lat]], {\it and references therein}.

\bibitem{haeg}  Ph.~Hagler {\it et al.}  [LHPC Collaborations],
  arXiv:0705.4295 [hep-lat].
 
\bibitem{Anselmino} M.~Anselmino, M.~Boglione, U.~D'Alesio, A.~Kotzinian, F.~Murgia, A.~Prokudin and C.~Turk,
  Phys.\ Rev.\  D {\bf 75}, 054032 (2007).

\bibitem{BrodskyLepage} G.~P.~Lepage and S.~J.~Brodsky,
  Phys.\ Rev.\ Lett.\  {\bf 43}, 545 (1979)

\bibitem{VdHLag} M.~Vanderhaeghen, M.~Guidal and J.~M.~Laget,
  Phys.\ Rev.\  C {\bf 57}, 1454 (1998); 
{\it ibid} Phys.\ Lett.\  B {\bf 400}, 6 (1997).
  
\bibitem{ong} P.~Kessler and S.~Ong,
   Phys.\ Rev.\  D {\bf 48}, 2974 (1993); S.~Ong,
  Phys.\ Rev.\  D {\bf 52}, 3111 (1995).  

\bibitem{StermanLi} H.~n.~Li and G.~Sterman,
  Nucl.\ Phys.\  B {\bf 381}, 129 (1992).
  
\bibitem{KroJak} R.~Jakob and P.~Kroll,
 Phys.\ Lett.\  B {\bf 315}, 463 (1993)
 
\bibitem{RCARnold} R.~C.~Arnold and M.~L.~Blackmon,
  Phys.\ Rev.\  {\bf 176}, 2082 (1968).
  
\bibitem{BalBra} P.~Ball and V.~M.~Braun,
  Nucl.\ Phys.\  B {\bf 543}, 201 (1999);
 {\it ibid}  arXiv:hep-ph/9808229.

\bibitem{BalBraTan} P.~Ball, V.~M.~Braun, Y.~Koike and K.~Tanaka,
  Nucl.\ Phys.\  B {\bf 529}, 323 (1998).
  
\bibitem{GolLiu} G.R. Goldstein and S.Liuti, {\it in preparation}.

\bibitem{DVCSA} C.~Munoz Camacho {\it et al.}  [Jefferson Lab Hall A Collaboration],
Phys.\ Rev.\ Lett.\  {\bf 97}, 262002 (2006)

\bibitem{DVCSB} F.~X.~Girod {\it et al.}  [CLAS Collaboration],
  Phys.\ Rev.\ Lett.\  {\bf 100}, 162002 (2008)
 
S \bibitem{HallA_proposal} C. Munoz Camacho {\it et al.}, Hall A Collaboration, 
Jlab PAC31 Proposal, ``Complete Separation
of Deeply Virtual Photon and $\pi^0$ Electroproduction Observables of Unpolarized Protons'' (2006).

\bibitem{demasi} R.~De Masi {\it et al.},
  Phys.\ Rev.\  C {\bf 77}, 042201 (2008).

\end{thebibliography}
\end{document}